\documentclass[aps,prb,floatfix,reprint]{revtex4-1}

\usepackage{graphicx}
\usepackage{subfigure}
\usepackage{amsfonts}
\usepackage{amsmath}
\usepackage{amssymb}

\begin{document}

\title{Theoretical Analysis of Metallic-Nanodimer Thermoplasmonics for Phototactic Nanoswimmers}

\author{Andr\'es I. Bertoni$^{1,3}$, Nicol\'as Passarelli$^{2,3}$, Ra\'ul Bustos-Mar\'un$^{2,*}$}
\affiliation{$^{1}$Instituto Interdisciplinario de Ciencias B\'asicas (ICB-CONICET), Universidad Nacional de Cuyo, Padre Jorge Contreras 1300, Mendoza 5502, Argentina.
$^{2}$Instituto de F\'isica Enrique Gaviola (IFEG-CONICET) and Facultad de Ciencias Qu\'imicas, Universidad Nacional de C\'ordoba, Ciudad Universitaria, C\'ordoba 5000, Argentina.
$^{3}$These authors contributed equally to this work.
}
\email{rbustos@famaf.unc.edu.ar}

\begin{abstract}
We assess the potentiality of several geometries of metallic nanodimers (one of the simplest thermoplasmonic systems) as candidates for active particles (nanoswimmers) propelled and controlled by light (phototaxis).
The studied nanodimers are formed by two spherical nanoparticles of gold, silver, or copper with radii ranging from 20 to 100 nm. Contrary to most proposals, which assume the asymmetry of the systems as a requirement for self-propulsion, our results show that nanodimers made of identical nanoparticles are excellent candidates for phototactic self-thermophoretic systems. Nonsymmetrical nanodimers, although having a tunable effective diffusion, possess much lower or zero average thermophoretic forces.
We show that the effective diffusion and the net thermophoretic force in both types of systems depend strongly on the wavelength of the incident light, which makes these properties highly tunable.
Our study may be useful for the design of simple-to-make but controllable self-propelled nanoparticles.
This can find numerous applications ranging from autonomous drug-carrying to controlling the self-assembly of complex nanomaterials.
\end{abstract}

\maketitle

\section{Introduction}

The design and control of nanomotors and molecular machines is a subject of great interest that has undergone enormous development in recent years.
Particularly appealing are the so-called ``nanoswimmers'', suspended nanoparticles (NPs) that navigate through a fluid thanks to mechanical forces that arise from inhomogeneities in their  surroundings ~\cite{golestanian2007,walther2008,wang2013,wu2016rev,lin2017,moran2017,xu2017,guix2018}.
Such inhomogeneities can be, e.g., gradients of concentration (diffusiophoresis)~\cite{howse2007,ruckner2007,lee2014,schattling2015,wang2015,qin2017} or temperature (thermophoresis)~\cite{jiang2010,buttinoni2013,baraban2013,kummel2013}.
Self-propulsion may also occur when nanoswimmers are able to produce their own local gradients.
However, in this case, it is more difficult to find the presence of taxis, or the guided motion towards or away from a stimulus source.
\begin{figure*}[ht]
   \centering
    \includegraphics[width=5 in]{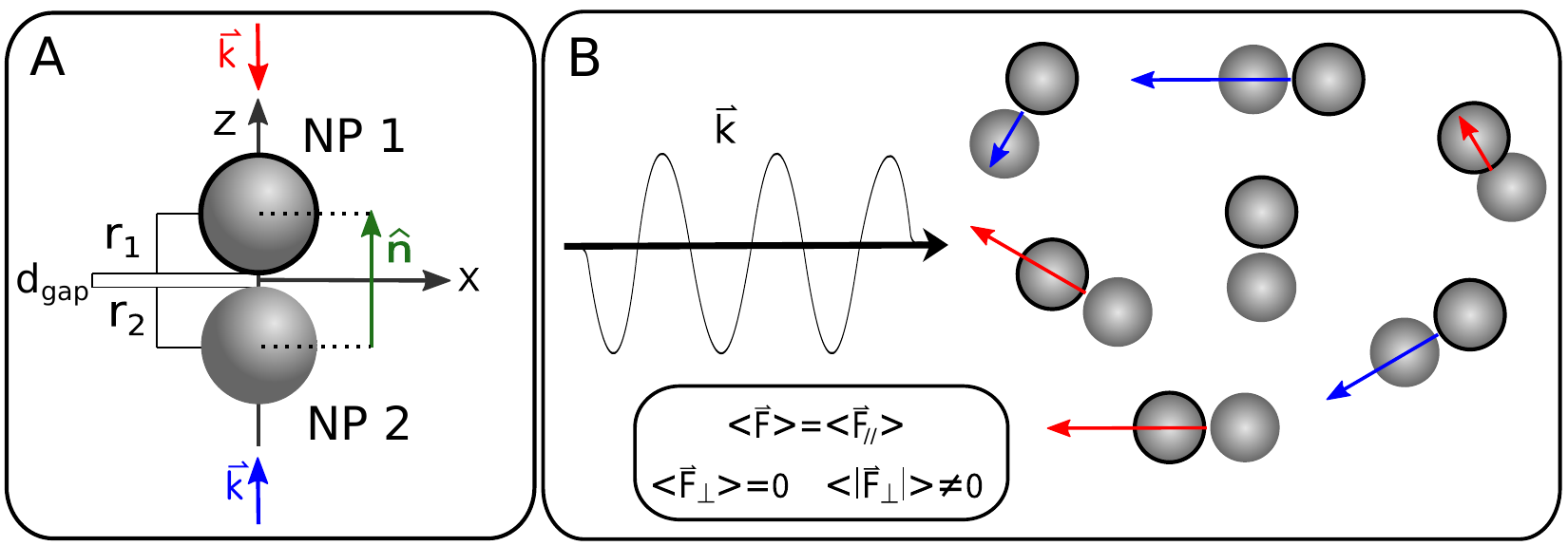}
    \caption{
    \textbf{(A) - }
     Diagram of a homodimer (two NPs of the same material) being illuminated by an electromagnetic field with wavevector $\vec{k}$. Depending on the relative orientation of the dimer and $\vec{k}$, NPs 1 and 2 are heated at different temperatures, which in turn causes a thermophoretic force.
    \textbf{(B) - } Scheme of an ensemble of thermophoretic nanoswimmers made of homodimers. Arrows depict the thermophoretic force ($\vec{F}$) expected for each orientation. Note that NP1 has a darker edge. }
    \label{fig:1}
\end{figure*}

As discussed in different reviews,\cite{wang2013,wang2015b,lin2017,moran2017,xu2017,guix2018} the range of applications of nanoswimmers (also called self-propelled particles, autonomous particles, or simply active matter) is vast. Some of the most studied are related to the possibility of increasing the reaction rates of chemical reactions by integrating the catalytic agents to autonomous particles.
This may lead, from an improvement in the performance of electrochemical sensors to a decrease in the degradation time of contaminants, just to mention some possibilities.
Furthermore, the use of nanoswimmers has also been extensively studied for different biological applications, as they may act as autonomous drug carriers.\cite{lin2017,guix2018}
Finally, collections of nanoswimmers are a form of synthetic active matter that has also been studied for controlling the self-assembly of complex nanomaterials.\cite{buttinoni2013,wang2015b}

Temperature gradients in the nanoscale give rise to multiple effects that are able to generate net forces over suspended NPs~\cite{jiang2010,golestanian2012}.
Those forces may arise from bubbles formation on the surface of the NPs~\cite{metwally2015,frenkel2017}, temperature-induced local phase segregation of mixed solvents~\cite{buttinoni2012}, kinetic activation of thermosensitive reactions~\cite{ma2016}, or simply because of the Soret effect of the solvent interacting with the surface of a particle having a temperature gradient
~\cite{jiang2010,yang2011,golestanian2012,baraban2013,buttinoni2013,kummel2013,yang2014,michaelides2015,ilic2016,wu2016,xuan2016}.

Typical nanoswimmers found in the literature are structures of different complexities but markedly asymmetric, e.g., Janus particles.
Mirror symmetry breaking may seem like a necessary condition for these systems, as there must be a preferential direction towards which to move.
However, because of retardation effects, when a light source illuminates a NP the electromagnetic fields around it are not necessarily mirror-symmetric along the direction of illumination.
In principle, this effect can then be used to obtain a directed phototactic motion of highly symmetrical nanoparticles.
In such a case, the symmetry breaking does not come from the structure itself but from its interaction with an electromagnetic field with a given direction.
Simpler and symmetrical nanoswimmers, in composition and structure, can be easier to produce and to test experimentally.
Furthermore, symmetries typically simplify and accelerate the theoretical calculations, allowing one to explore a wider range of parameters.
It is interesting then, to study how simple and symmetrical a thermophoretic nanoswimmer can be, while keeping its ability to generate an externally controllable thermophoretic force.

One of the most simple nanostructure to make in a laboratory is a dimer made of spherical metallic NPs.
They can be synthesized with different techniques that allow controlling sizes and separation between them~\cite{lan2013,sheikholeslami2010,sun2015,zheng2015}.
Moreover, some of these techniques can be extended to non-noble plasmonic metals~\cite{jiang2019,sun2015,garcia2014}.
The interaction with the light of nanodimers has an analytic solution and there are simple methods to estimate the temperature of each NP~\cite{baffou2010}.
Additionally, a general scaling analysis shows that nanoswimmers should be more efficient energetically~\cite{sabass2010} than microswimmers.
However, in this case, tighter control is desirable as Brownian forces become more important in this size scale.
Here, we study metallic nanodimers as light-induced thermophoretic nanoswimmers, see Fig. \ref{fig:1}.
As we will show, even symmetrical nanodimers, made of the same material and radii, are able to produce the asymmetrical temperature profiles required to generate thermophoretic forces.
Furthermore, those systems show a phototactic character and a strong dependence on the frequency, which can be important to externally control the nanoswimmers.

This work is organized as follows.
In section ``Theory'' we describe the theory used to calculate the light-induced temperature difference between the NPs of the nanodimer, and how this quantity is connected with the different average thermophoretic forces to which an ensemble of randomly oriented nanodimers is subject.
In section ``Employed Methodology'' we describe in more detail the studied system and the numerical calculations done to obtain the wavelength-dependent average thermophoretic forces.
In section ``Results'' we present the main results of this work while in section ``Conclusions'' we discuss their importance.

\section{Theory\label{sec:Theory}}

The basis for the motion of thermophoretic nanoswimmers is the mechanical force that arises from the temperature differences around the NPs.
To achieve a temperature gradient, a self-propelled particle should act as its own inhomogeneous source of heat.
This implies, for light-driven devices, that the particles must have a differential absorption rate along its structure.
The different parts of a NP may interact electromagnetically in nonintuitive ways typically imposing the use of numerical methods.
Therefore, for finding an estimation of the forces that move the nanoswimmers, it is needed to explicitly solve two problems, namely the light scattering and the photothermal conversion of the systems.

The full analytical solution for the light scattering of an arbitrary array of spheres is widely known, and there are several computational implementations of it~\cite{xu1997,pellegrini2007}.
The solution is built from an expansion in spherical harmonics of the incident, internal and scattered fields.
The coefficients of the expansion are then found by taking the boundary conditions resulting from centering the coordinates on each sphere and projecting the incident and the scattered fields by using the translational theorem.
In this formalism, the total absorption cross-section $C_\mathrm{abs}$ can be written as a sum over individual cross-sections $C_\mathrm{abs}^{(i)}$~\cite{xu1997,pellegrini2007},
\begin{equation}
C_\mathrm{abs}=\sum_{i=1}C_\mathrm{abs}^{(i)}.\label{eq:gmm}
\end{equation}
However, care must be taken as the terms $C_\mathrm{abs}^{(i)}$ does not correlate properly with the power absorbed by the individual NPs, $W_\mathrm{abs}^{(i)}$, which is formally given by~\cite{bohren2008}
\begin{equation}
W_\mathrm{abs}^{(i)}=\frac{\omega\varepsilon_{0}\varepsilon''}{2}\int_{V_i}\left|E(\mathbf{r})\right|^{2}d\mathbf{r}.\label{eq:abs-each}
\end{equation}
Here, $\omega$ is the angular frequency, $\varepsilon_{0}$ is the vacuum permittivity, $\varepsilon''$ is the imaginary part of the dielectric constant of the material, and $E(\mathbf{r})$ is the electric field inside the volume $V_i$ of each $i$-th sphere.
Once all $W_\mathrm{abs}^{(i)}$ terms are known, it is possible to estimate $\delta T^{(i)}$, which is the steady-state temperature difference of NP $i$ with respect to the surrounding nonabsorbent medium.
Here, we use the Green's function approach proposed by Baffou et.al~\cite{baffou2010} to calculate $\delta T^{(i)}$,
\begin{equation}
\delta T^{(i)}=\frac{1}{4\pi\kappa}\left(\frac{W_\mathrm{abs}^{(i)}}{a^{(i)}}+\frac{W_\mathrm{abs}^{(j)}}{d}\right),
\label{eq:GF_DT}
\end{equation}
where $\kappa$ is the thermal conductivity of the surrounding medium, water in our case, $a^{(i)}$ is the radius of the $i$-th NP, and $d$ is the distance from center to center between the two NPs of the dimer.
This approach assumes the temperature on each NP as a constants, which should be a good approximation for metallic NPs,  and considers the NPs as point-like heat sources.
Although deviations may appear with respect to more realistic calculations, the purpose of the present work is to address the general behavior of the studied system and the physics behind it, not the exact value of the thermophoretic forces.
For the same reason, we do not perform hydrodynamic or molecular dynamic simulations~\cite{yang2011,yang2014} to assess the exact value of the thermophoretic forces $\mathbf{F}$.
Instead, we take only the leading order of the expansion of $\mathbf{F}$ in terms of the temperature difference between the NPs, $\Delta T=\delta T^{(1)} - \delta T^{(2)}$, where we assume the force points from the center of one NP to the center of the other one, see Fig.~\ref{fig:1}-(A). In such a case, the thermophoretic force $\mathbf{F}$ can be written as~\cite{yang2011,lin2017}
\begin{equation}
\mathbf{F}=-\alpha_f k_B \Delta T \mathbf{\hat{n}}\label{eq:force}
\end{equation}
where $\hat{\mathbf{n}}$ is the unit vector that points from the center of the sphere $(2)$ to the center
of sphere $(1)$, $\alpha_f$ is the thermal diffusion factor and $k_B$ is the Boltzmann constant.
Note that, not only the magnitude but also the sense of the vector $\mathbf{F}$ will be determined by the value of $\alpha_f$. Numerical simulations show that the thermal diffusion factor can be positive or negative depending on the nature of the interaction between the NPs and the surrounding medium~\cite{yang2011,yang2014}.

The presence of a thermophoretic force induces the movement of the nanodimer in the direction of its bond. Typically after a short time, friction forces balance other forces and then the system reaches a steady-state with average velocity $V$ such that~\cite{jiang2010,yang2011,lin2017}
\begin{equation}
 V=\frac{1}{\gamma}|\mathbf{F}| , \label{eq:V}
\end{equation}
where $\gamma$ is the friction coefficient.
At micro and nanoscale, the interaction with the environment not only moves the center of mass of the NPs but also randomly changes their orientation. 
The change in the orientation of particles has two effects.
On the one hand, it typically averages the net velocity of the NPs to zero, but on the other hand, it gives rise to an enhanced diffusive behavior caused by $V \neq 0$.
This behavior is usually described by an effective diffusion coefficient $D_\mathrm{eff}$ given by~\cite{golestanian2007,wang2015},
\begin{equation}
D_\mathrm{eff} = D_{T}+\frac{V^{2}}{4D_{R}}, \label{eq:Deff_Janus}
\end{equation}
where $D_{T}$ and $D_{R}$ are respectively the Brownian translational and rotational diffusion coefficients. Note that $D_{T}$ is the diffusion coefficient of the particle at $V=0$ (passive particle).
Therefore, even if an ensemble of particles has zero net force, the diffusive behavior of the system can be anyway affected by thermophoretic forces as this effect depends on the modulus of force only.
For that reason, it is important to not only study the mean thermophoretic force $\left < \mathbf{F} \right >$, but also the average of its modulus $\left < |\mathbf{F}| \right >$.

When an ensemble of nanoparticles in solution is illuminated from a given direction, each particle senses light from a different (random) direction, see Fig.~\ref{fig:1}-(B). In this case, it becomes relevant to study the average force over the nanodimers $\left\langle \mathbf{F}\right\rangle$,
\begin{eqnarray}
\left\langle \mathbf{F}\right\rangle &=&
\mathbf{\hat{i}}\left\langle F_{x}\right\rangle +\mathbf{\hat{j}}\left\langle F_{y}\right\rangle +\mathbf{\hat{k}}\left\langle F_{z}\right\rangle, 
\end{eqnarray}
where the $\left\langle F_{\alpha} \right\rangle$s are the components of the average force along the different axes $\alpha$, and $\hat{i}$, $\hat{j}$ and $\hat{k}$ are the unit vectors that set the direction of the axes. Let us take the spherical coordinate system where the nanodimer direction $\hat{n}$ matches the $z$ axis, $\theta$ is the polar angle between the wavevector $k$ of the incident light and $\hat{n}$ ($\theta=\{ 0,\pi \}$), while $\phi$ is the azimuthal angle setting the projection of the wavevector on the $x-y$ plane ($\phi=\{ 0,2 \pi \}$).
Now, because of the cylindrical geometry, only the angle $\theta$ and the polarization of light are relevant quantities of the treated system, see Fig.~\ref{fig:1}-(A). Therefore, the temperature gradient can only be a function of $\theta$ and the polarization angle of light. Then, the components of the force are
\begin{eqnarray}
F_{x} & = & -\alpha_f k_B \sin (\theta) \cos (\phi) \Delta T\left(\theta,\mathrm{pol}\right), \notag \\
F_{y} & = & -\alpha_f k_B\sin (\theta) \sin (\phi) \Delta T\left(\theta,\mathrm{pol}\right), \notag \\
F_{z} & = & -\alpha_f k_B \cos (\theta) \Delta T\left(\theta,\mathrm{pol}\right), 
\end{eqnarray}
while the components of the average force are
\begin{eqnarray}
\left\langle F_{x}\right\rangle & = & -\frac{\alpha_f k_B }{4 \pi}
\intop_{0}^{2\pi} \cos \phi \mathrm{d}\phi \intop_{0}^{\pi}
\overline{\Delta T}
\sin^{2}\theta \mathrm{d}\theta,  \notag \\
\left\langle F_{y}\right\rangle & = & -\frac{\alpha_f k_B }{4 \pi}
\intop_{0}^{2\pi} \sin \phi \mathrm{d}\phi \intop_{0}^{\pi}
\overline{\Delta T}
\sin^{2}\theta \mathrm{d}\theta , \notag \\
\left\langle F_{z}\right\rangle & = & -\frac{\alpha_f k_B }{2}
\intop_{0}^{\pi}
\overline{\Delta T}
\cos\theta\sin\theta\mathrm{d}\theta.
\label{eq:F_parall}
\end{eqnarray}
where $\overline{\Delta T}$ is the value of $\Delta T$ averaged over the polarization angle of light.
Clearly the mean thermophoretic force can only have a parallel component, $ \langle \mathbf{F}_{//} \rangle = \mathbf{\hat{k}} \left\langle F_{z}\right\rangle$, since its perpendicular component, $\left\langle \mathbf{F}{\perp} \right\rangle =   \mathbf{\hat{i}} \left\langle F_{x} \right\rangle + \mathbf{\hat{j}} \left\langle F_{y}\right\rangle$, is always zero by symmetry.

Note that, although the average perpendicular force $\langle \mathbf{F}_{\perp} \rangle$ is null, its average modulus is not in general, 
\begin{equation}
\langle |F_{\perp}| \rangle = \frac{|\alpha_f| k_B }{\pi}
\intop_{0}^{\pi} |\overline{\Delta T}|  \sin^{2}\theta \mathrm{d}\theta . 
\label{eq:F_perp_mod}
\end{equation}
where $\langle |F_{\perp}| \rangle = \langle |F_x| \rangle = \langle |F_y | \rangle$.
As discussed, the modulus of the thermophoretic force is an important quantity as it may give rise to enhanced diffusive behaviors. The possibility of externally controlling the diffusion of the system is one of the interesting aspects to be studied, but this becomes even more appealing if the enhanced diffusive behavior is anisotropic, i.e. if $\left\langle |F_{\perp}| \right\rangle \neq \left\langle | F_{//} | \right\rangle$ where
\begin{equation}
\langle |F_{//}| \rangle  = \frac{|\alpha_f| k_B }{2}
\intop_{0}^{\pi} |\overline{\Delta T}| \left |\cos\theta \right |  \sin\theta \mathrm{d}\theta . 
\label{eq:F_parall_mod}
\end{equation}
This could lead to a way of controlling the diffusion coefficient of NPs in a particular direction, a subject that can be of interest in the field of active-particle dynamics~\cite{bechinger2016}.

It is worth mentioning, that in this work we are only interested in describing general properties of the thermoplasmonic forces acting on illuminated metallic nanodimers. For that reason, we do not consider explicitly the constant $\alpha_f$ and thus, throughout the manuscript, forces are shown in arbitrary units and figures are discussed only in relative terms. Despite this, in the appendix ``Estimation of thermophoretic forces and velocities'' we provide a rough estimation of the average thermophoretic force and the steady-state velocity of a particular example.

\section{Employed Methodology\label{sec:methods}}
\subsection{System Description}

Fig. 1 shows a general diagram of the system treated here, two spherical particles with radii $r_1$ and $r_2$ separated by a gap $d_{gap}$. We numerically explore different geometries and materials of the dimers which we label as ``homodimer/heterodimer'' and ``symmetrical/asymmetrical''.
The terms ``homodimer'' and ``heterodimer'' stand for dimerical systems where the NPs are of the same or different materials respectively. The terms ``symmetrical'' and ``asymmetrical'' distinguish NPs of the same ($r_1=r_2$) or different ($r_1 \neq r_2$) radii respectively.

Notice that, independently of the size or material of the particles, the system has always cylindrical symmetry. 
Thus, only three parameters need to be varied for a fixed geometry: the angle $\theta$, the polarization of the incident light and its wavelength $\lambda$.
The wavelength was varied in the $300-800\,nm$ interval. Such interval corresponds to the extended visible range. Because we intended an analysis independent of the polarization of the incident light, we always average the results obtained from the two orthogonal linear polarizations.

The model used to describe the system takes into account a nonabsorbent medium in which both particles are immersed. In our calculations, we used the refractive index of water for the medium, $n=1.335238$. This value was taken from the literature~\cite{hale1973} and corresponds to the average refractive index of experimental measurements of pure water between $300\,nm$ and $800\,nm$ with a temperature of $298\,K$ and a pressure of $1\,bar$.	

\subsection{General Procedure}
\begin{figure}[ht]
    \centering
    \includegraphics[width=2.8 in]{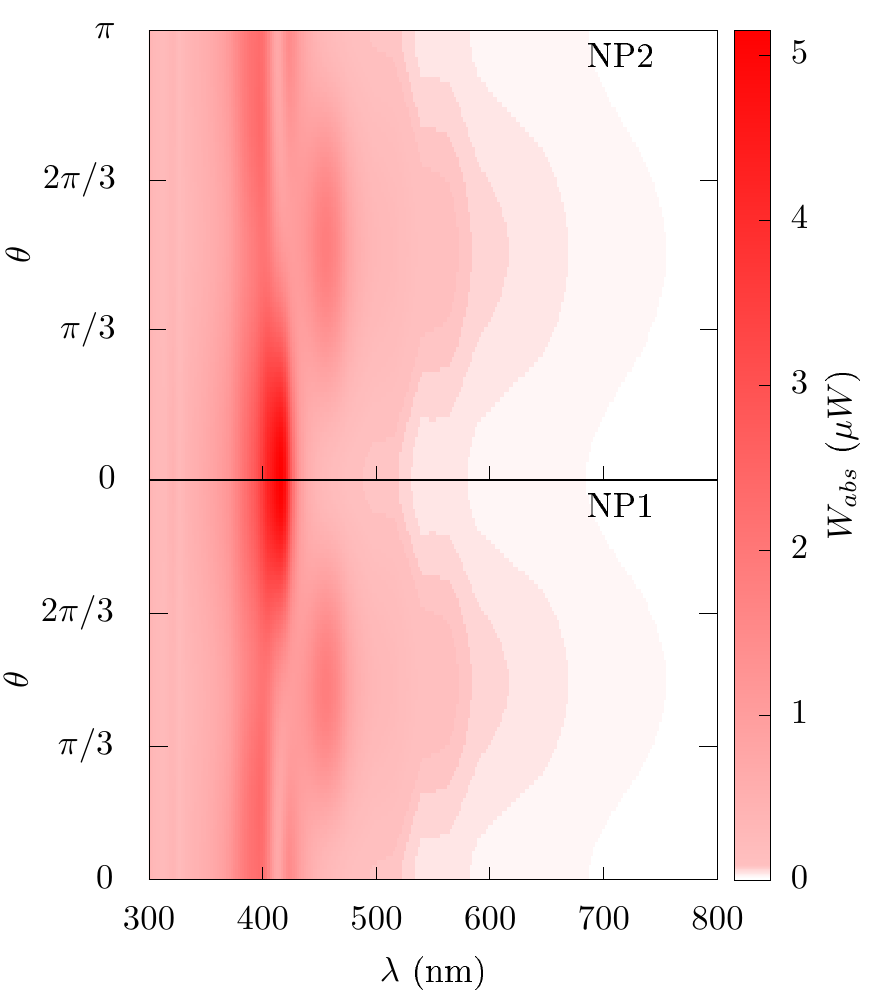}
    \caption{Absorption powers ($W_\mathrm{abs}$) of each NP of an Ag-Ag dimer in aqueous solution as a function of the wavelength $\lambda$ and the polar angle $\theta$ of the incident wave. The separation gap is $5\,nm$ and $r_1 = r_2 = 20\,nm$. The results shown correspond to the average of the different polarizations of the incident light.}
    \label{fig:2}
\end{figure}

\begin{figure}
    \centering
	\includegraphics[width=2.8 in, trim={0.0in 0.0in 0.0in 0.1in},clip]{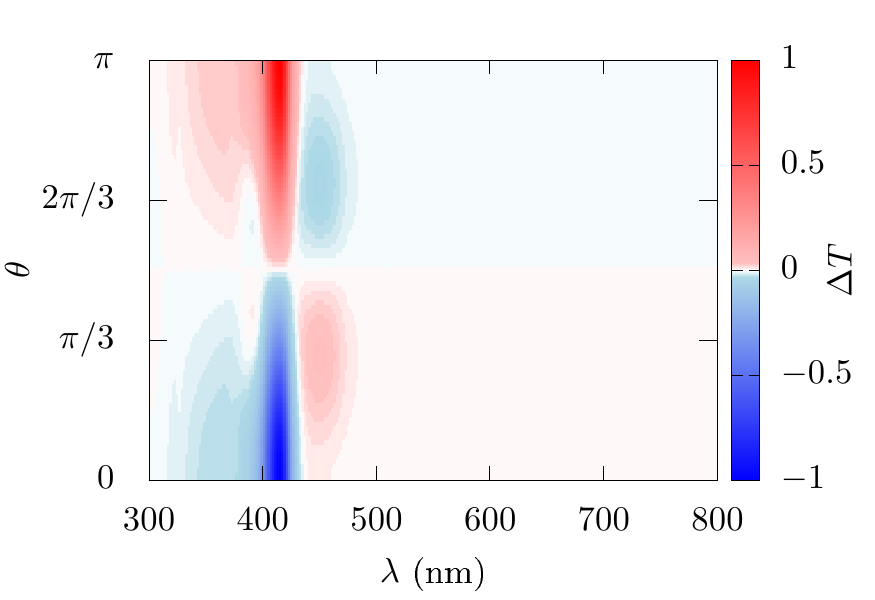}	
    \caption{Normalized temperature difference between the NPs of an Ag-Ag dimer as a function of the wavelength $\lambda$ and the polar angle $\theta$ of the incident light. The parameters of the dimer are the same as those of Fig.~\ref{fig:2}}
    \label{fig:3}
\end{figure}

\begin{figure}[ht]
	\centering
	\includegraphics[width=3.0 in, trim={0.0in 0.0in 0.15in 0.08in},clip]{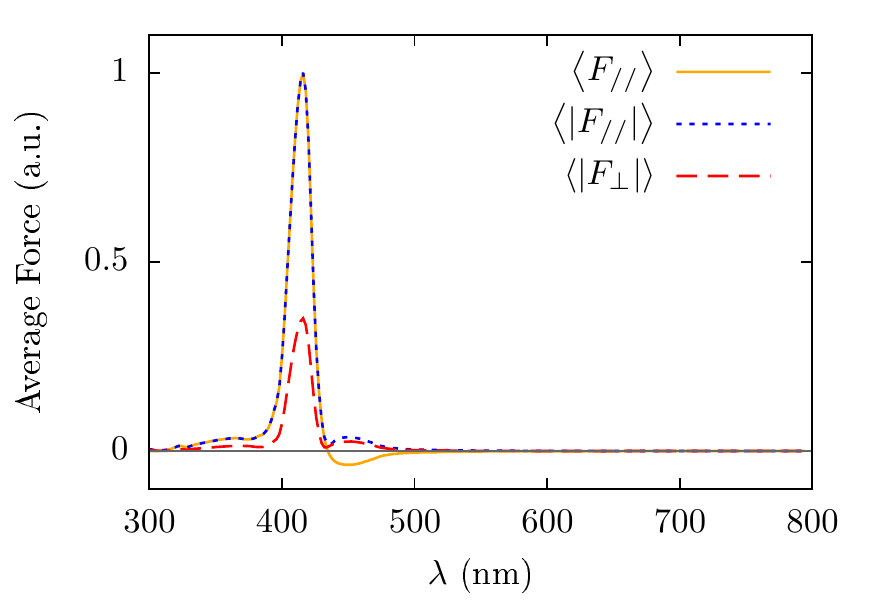}
	\caption{Different components of the average thermophoretic force of Ag-Ag dimers as a function of $\lambda$. The parameters of the dimer are the same as those of Fig.~\ref{fig:2}.
	The forces are normalized with respect to the maximum value of $\langle F_{//} \rangle/(|\alpha_f| k_B)$.}
	\label{fig:4}
\end{figure}

This section details the general procedure employed to obtain the force profiles of the studied nanodimers.
We wrote a code for the systematic analysis of the dimer which automates all the steps described below. The code used the open source subroutines developed by Pellegrini and Mattei~\cite{pellegrini2007} for the calculations of near fields within the theoretical framework of the Generalized Multiparticle Mie theory (GMM).
To better illustrate all the steps involved in the calculations, we take the silver (Ag) homodimer as an example in Figs.~\ref{fig:2}-\ref{fig:4}.

\textit{Outlining the system.} 
We first set the values of all the parameters that define the dimer: both NPs radii (\,$r_1$ and $r_2$\,), the distance between them ($d_{gap}$) and its constituent materials. The latter relies on literature's experimental measurements of the complex electrical permittivity constant ($\epsilon$) of the material, determined for a range of incident wavelengths. For the case of the main example, it consists of a Ag homodimer with $r_1 = r_2 = 20\,nm$ separated by a gap $d_{gap} = 5\,nm$. The values of the complex electrical permittivity ($\epsilon$) of Ag were taken from the experiments done by Ferreiro et al~\cite{ferreiro2009}. The optical constants for the other materials (Cu and Au) also come from experimental measurements, see Refs.~\cite{johnson1972,palik1998}.

The exploration of the polar angle ($\theta=\{0,\pi\}$) was made by steps of $\pi/45$. Each configuration for a given $\theta$ was evaluated for an incident light of wavelengths between $300\,nm$ and $800\,nm$, by steps of $2\,nm$.
This exploration defines a grid of $91 \times 251$ points, where each point is a different calculation of the same dimeric system but for different incident light.
Said grid was executed twice, to take into account the two orthogonal polarizations of the incident light.

\textit{Power absorbed by each NP.}
Each calculation of the exploration grid consisted in the integration of the near electric field to evaluate $W_{\mathrm{abs}}^{(i)}$, see Eq.~\ref{eq:abs-each}. The operation has to be made separately over the volume of each sphere. Those results were then averaged between the two orthogonal linear polarizations. That allowed the generation of 2D maps of the integrated quantities as a function of wavelength $\lambda$ and polar angle $\theta$ of the incident light, see Fig.~\ref{fig:2}.

Although thermophoretic forces will be shown afterward in arbitrary units, the plot shown here of $W_\mathrm{abs}^{(i)}$ was done by assuming an irradiance equal to $1\,.\,10^9\,W/m^2$. 
This value was taken from Ref.~\cite{baffou2010ACS} and is within the order of the values used in Ref.~\cite{jiang2010}.

\textit{Temperature differences.} 
As explained in section ``Theory'', we use the Green's function approach for calculating the steady-state temperature of each NP~\cite{baffou2010} with respect to its surroundings, Eq.~\ref{eq:GF_DT}.
Figure~\ref{fig:3} shows the temperature difference as a function of the wavelength of the incident field, $\lambda$, and its angle of incidence ($\theta$). In the plot, this quantity was normalized to its maximum values.

The values of $\Delta T$ were calculated using a somewhat simplistic theory, Eq. \ref{eq:GF_DT}, that although allowed us to explore a wide range of geometries and materials of the dimers, it is expected to overestimate $\Delta T$ due to the underestimation of interparticle thermal interactions.
For this reason, we only assume as valid the qualitative behavior of $\Delta T$, and the quantities derived from it.
In this regard, in Ref. ~\cite{baffou2010ACS}, the authors calculated $\Delta T$ as a function of $\lambda$ and $\theta$ for a symmetrical gold homodimer by using a more sophisticated numerical procedure. They found, among other things, that the relative temperature difference closely follows the relative power absorption difference. Note that for symmetrical dimers, $\Delta T \propto (W_{abs-1}-W_{abs-2})$ according to Eq. \ref{eq:GF_DT}. Therefore, even though the absolute values of $\Delta T$ may be overestimated, at least in this case is clear that Eq. \ref{eq:GF_DT} provides a good estimation of the qualitative behavior of $\Delta T$.

\textit{Average thermophoretic forces.}
As discussed in section ``Theory'', the absolute value of the computed temperature difference ($ | \Delta T | $) is proportional, within the approximation used, to the modulus of thermophoretic force acting on the dimer for fixed values of $\lambda$ and $\theta$.
The direction of the force is given by the versor ($\mathbf{\hat{n}}$) pointing from the center of one NP to the center of the other.

From the previously computed maps of $\Delta T (\theta,\lambda)$, and using Eqs.~\ref{eq:F_parall},~\ref{eq:F_perp_mod} and~\ref{eq:F_parall_mod}, we calculate the mean parallel component of the thermophoretic force ($\langle F_{//} \rangle$), its average modulus ($\langle |F_{//}| \rangle$),  and the average modulus of the perpendicular component of the thermophoretic force ($\langle |F_{\perp}| \rangle$). In Fig.~\ref{fig:4} we show a typical plot of $\langle F_{//} \rangle$, $\langle |F_{//}| \rangle$ and $\langle |F_{\perp}| \rangle$ as function of the wavelength  $\lambda$ of the incident light. 

\section{Results\label{sec:results}}
\begin{figure}[ht]
	\centering
	\includegraphics[width=2.8 in, trim={0.0in 0.2in 0.15in 0.08in},clip]{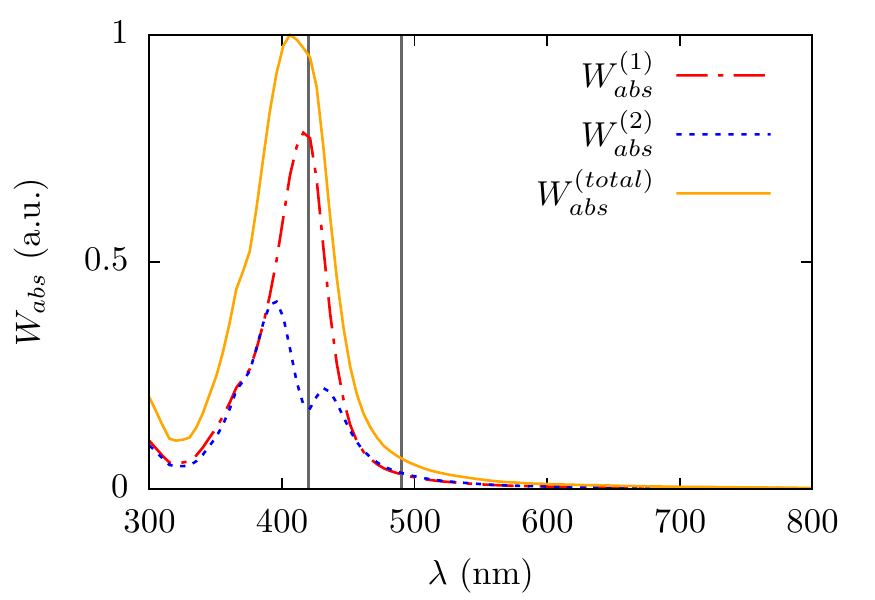}\\
	\includegraphics[width=2.8 in, trim={0.0in 0.0in 0.15in 0.08in},clip]{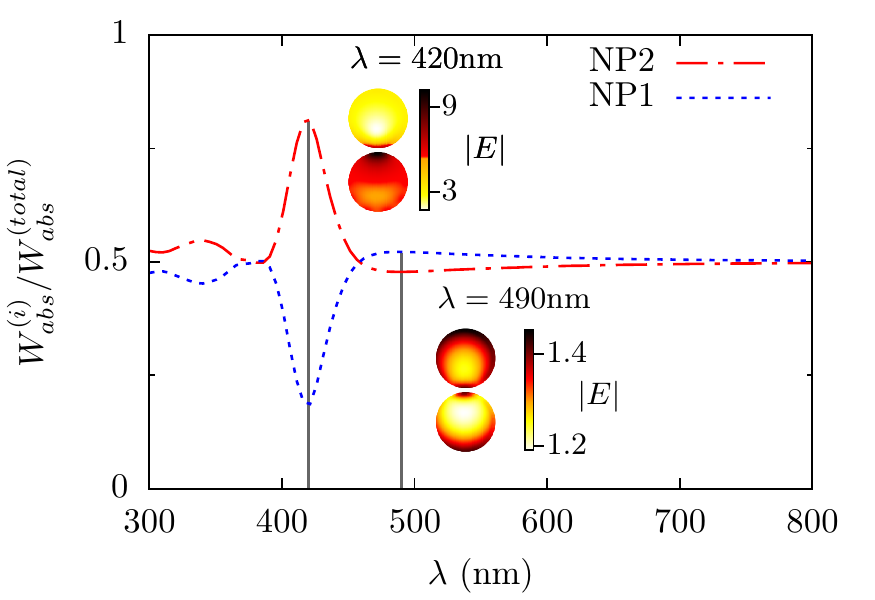}\\	
	\caption{Absorption power of each NP ($W_\mathrm{abs}^{(i)}$) and the total absorption power ($W_\mathrm{abs}^\mathrm{total}=W_\mathrm{abs}^{(1)}+W_\mathrm{abs}^{(2)}$) of an Ag-Ag dimer being illuminated with light of wavelength $\lambda$ and $\theta=0$.
	The parameters of the dimer are the same as those of Fig.~\ref{fig:2}.
	In the upper panel, $W_\mathrm{abs}$ is normalized with respect to its maximum value. The insets in the lower panel show the intensity of the electric fields inside the NPs (normalized with respect to the incident field) for two values of $\lambda$.}
	\label{fig:5}
\end{figure}

The first goal of this work is to assess the potentiality of metallic nanodimers as thermophoretic nanoswimmers.
Fig. \ref{fig:4} shows that some metallic nanodimers present strong phototaxis ($\langle F_{//} \rangle \neq 0 $). Therefore, illuminated nanodimers such as the AgAg(20,20)
(for silver-silver symmetrical homodimer of $r_1=r_2=20\ nm$) will not only present thermophoretic forces but they will also exhibit net velocity parallel to the direction of illumination.

One surprising feature of the dependence of thermophoretic forces with $\lambda$ is the change of sign of the average force $\langle F_{//} \rangle$, see Fig. \ref{fig:4} at $\lambda \approx 450 nm$.
To further explore this point, in Figure.~\ref{fig:5} we plot the total power absorbed by the AgAg(20,20) nanodimer, as well as the power absorbed by each NP as a function of $\lambda$.
In the lower panel of the same figure, we showed the relative contribution of each NP to the total absorbed power and, as insets, the electric fields inside the NPs for the two wavelengths where $\langle F_{//} \rangle$ has its maxima with different signs. Clearly, the change of sign of the temperature difference is a direct consequence of the concentration of the electromagnetic fields over one or the other NP.
The same kind of phenomena has been observed before for short chains of NPs.~\cite{2005hernandez,2007dewaele,2008malyshev}
Retardation effects are the main responsible for this behavior.
Because of them, the external field and the fields emitted by each NP interfere constructively or destructively over each NP depending on $\lambda$.
This, in turn, causes that the NP closest to the light source (or the other one depending on $\lambda$) heat the most.

One important aspect of the system studied is the possibility of externally controlling the direction of motion of the active particles. In this respect, the increase of the effective diffusion coefficient due to thermophoretic forces, see Eqs. \ref{eq:V} and \ref{eq:Deff_Janus}, is an undesirable effect. As shown in Fig. \ref{fig:4} for the case of AgAg(20,20), the increase of the net thermophoretic force $\langle F_{//} \rangle$ is usually accompanied by an increase of the perpendicular diffusion, related with $\langle | F_{\perp}| \rangle$. Of all the geometries studied, the AgAg(20,20) nanodimer shows some of the largest value of $\langle F_{//} \rangle$, or more precisely some the largest value of $\langle F_{//} \rangle/|\alpha_f| k_B$, and one of the smallest value of $\langle | F_{\perp}| \rangle$ relative to $\langle F_{//} \rangle$.
Moreover, the dependency of $\langle F_{//} \rangle$ with $\lambda$, or the force spectrum, shows a sharp peak which is also very sensitive to the geometry of the dimer.
This is an additional advantage since this feature can be used to selectively control a given type of nanodimer without affecting much the others.
Therefore, although more studies are necessary, our results suggest that the symmetrical homodimer made of silver NPs with a radius of $20 nm$ seems like one of the most promising candidates for a phototactic thermophoretic nanoswimmer (for radii up to $40\, nm$ the system behaves similarly).
In the following, we will discuss some of the general characteristics of other nanodimers, similar to AgAg(20,20) but made with larger NPs, NPs of different materials (symmetrical heterodimers), or NPs of different radii (asymmetrical homodimers).

\subsection{Homodimers and heterodimers}

\begin{figure}[ht!]
    \centering    
     \includegraphics[width=3.3 in]{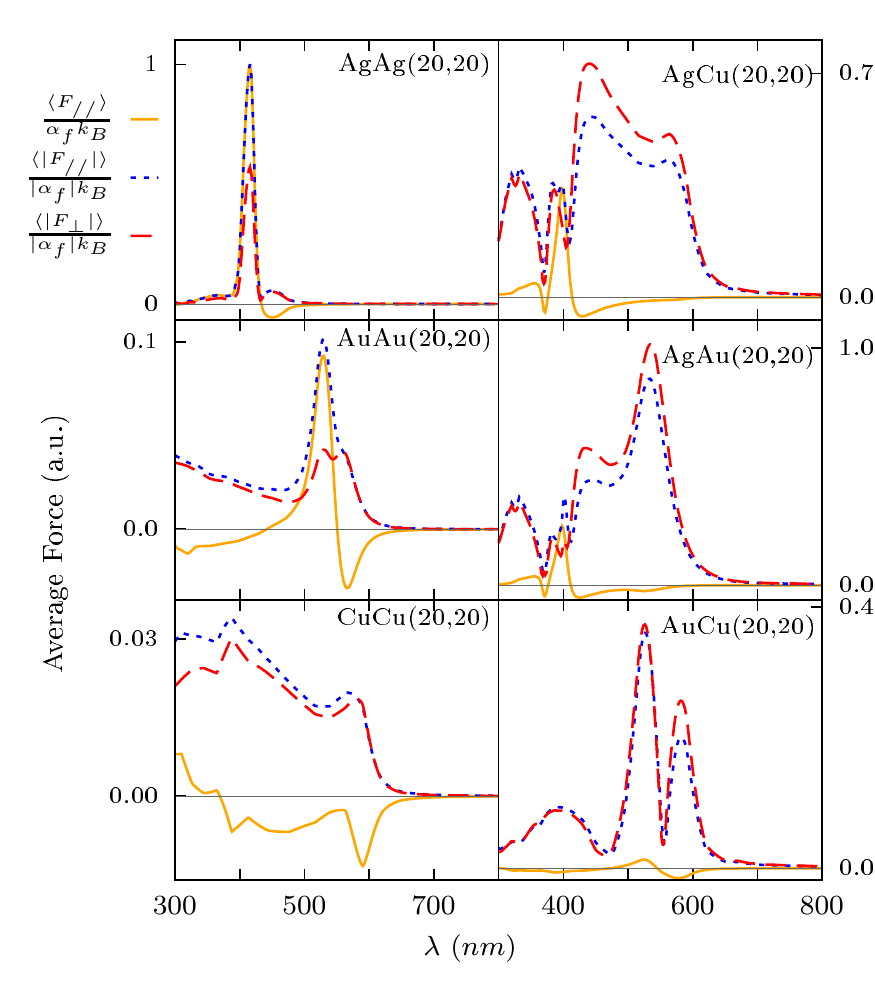} 
    \caption{Force spectra for symmetrical homodimers (\textit{left}) and symmetrical heterodimers (\textit{right}) made of NPs of silver (Ag), gold (Au) and copper (Cu), with $r_1=r_2=20\,nm$ and $d_{gap}=5\,nm$. We used the maximum of the force spectrum of AgAg(20,20) to normalize the figures.}
  \label{fig:6}
\end{figure}
We first study some representative examples of the effect of the material of the NPs over thermophoretic forces and the differences between homo- and heterodimers. Fig. \ref{fig:6} shows the force spectra of symmetrical homo- and heterodimers made of different combinations of silver (Ag), gold (Au), or copper (Cu) NPs.

Although the value of $\alpha_f$ may change with the material of the NPs, our results suggest that 
the systems that comparatively show the largest average temperature difference ($\langle F_{//} \rangle/|\alpha_f| k_B$), and therefore the largest thermophoretic forces, are those made of two NPs of the same material. Also noteworthy is that the nanodimer's force profile appears to be highly sensitive to the material.
This may have important applications for developing new separation techniques, for example.

Unlike homodimers, the force spectra of heterodimers show regions where the average force ($\langle F_{//} \rangle$) is negligible while the average modulus of the parallel and perpendicular components of the force ($\langle |F_{//}| \rangle\,$ and $\langle |F_{\perp}| \rangle$) are still important.
This will cause the ensemble of nanodimers to spread over all directions, without exhibiting a net displacement.
Although this fact makes them suboptimal phototactic nanoswimmers, they still present thermophoretic forces ($\langle |F_{\perp}| \rangle \neq 0$ and $\langle |F_{//}| \rangle \neq 0$) which should alter their average velocities and hence the effective diffusion coefficients of the particles.
It is interesting that, within certain spectral regions, these systems should exhibit anisotropic diffusion ($\langle |F_{\perp}| \rangle \neq \langle |F_{//}| \rangle$).
The above facts imply that heterodimers can still be useful for many purposes, e.g., to study in a controllable manner the effect of the effective diffusion coefficient of active particles on some property.
Note that, according to our results, the effective diffusion coefficient of an illuminated nanodimer should depend on the laser's intensity and wavelength, besides the size and composition of the NPs.

Another interesting aspect of heterodimers comes from taking into consideration that the region of space illuminated by the laser is finite (the ``\textit{light spot}''). Then, nanoswimmers reached by the beam of light will have a greater effective diffusion coefficient than those in the surrounding dark regions. After a while, that would entail a depletion of nanoswimmers in the light spot; thus, in the presence of inhomogeneous illumination, concentration gradients should form.

\subsection{Symmetrical and asymmetrical homodimers}

\begin{figure}[ht!]
    \includegraphics[width=3.3 in]{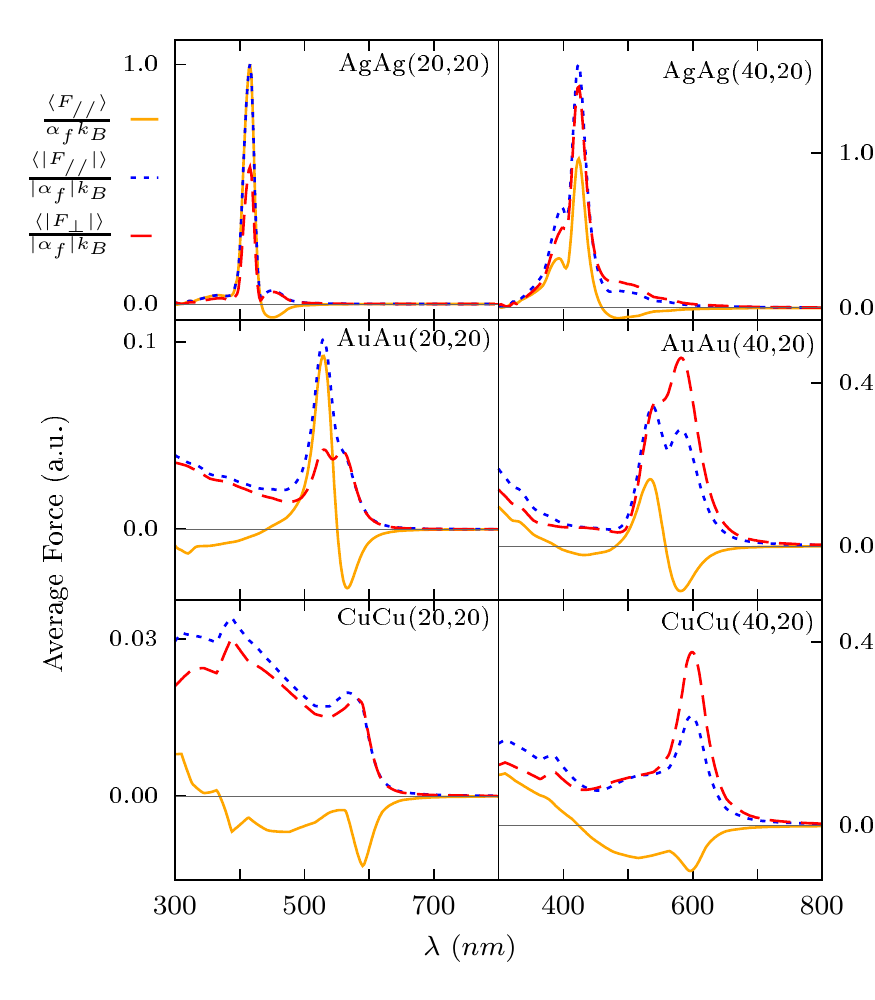} 
    \caption{Force spectra of Ag, Au and Cu symmetrical homodimers \textit{(left)} with $r_1=r_2=20\,nm$, and force spectra of asymmetrical homodimers \textit{(right)} with $r_1=40\,nm$, $r_2=20\,nm$. The gap is $5\,nm$ and we used the maximum of the force spectrum of AgAg(20,20) to normalize the figures.}
    \label{fig:7}    
\end{figure}

Now let us analyze the effect of the size difference between the NPs of the dimer.
Fig. \ref{fig:7} compares the force spectra of Ag, Au and Cu symmetrical ($r_1 = r_2$) homodimers (same material) with those of asymmetrical ($r_1 \neq r_2$) homodimers. 
As can be seen in the figure, asymmetrical systems show in general similar average forces ($\langle F_{//} \rangle/\alpha_f k_B$) with respect to their symmetrical counterparts but with much larger values of
$\langle | F_{//} | \rangle/\alpha_f k_B$ and $\langle | F_{\perp} | \rangle/\alpha_f k_B$, i.e., ensembles of asymmetrical nanodimers should have a much larger dispersion.
For that reason, symmetrical homodimers seem to be better candidates for phototactic nanoswimmers with respect to asymmetrical homodimers.

\subsection{Symmetrical homodimers of increasing radii}

As we have shown, symmetrical homodimers seem like the best candidates for thermophoretic nanoswimmers controlled by light. The question that arises now is what is the optimal size of the nanodimers.
To address this question, we systematically changed the radius of the NPs from $20\,nm$ to $100\,nm$ by steps of $10\, nm$ (not all the force spectra are shown in the figures). We purposely exclude Cu homodimers from the analysis as they possess, by far, the smallest average temperature differences, up to 2 orders of magnitude smaller compared with silver nanodimers.
The lower limit was chosen according to the existing experimental difficulty for the synthesis of stable and monodispersed metallic NPs of radii lower than $10\,nm$ or $20\,nm$. In addition, the absorption spectra of Ag and Au NPs in a colloidal solution do not show significant changes for radii beneath approximately $20\,nm$~\cite{agnihotri2014,link1999}. The upper limit of $100\,nm$ also corresponds to an experimental limitation: large NPs in solution tend to coagulate and precipitate as aggregates.
\begin{figure}[ht!]
	\includegraphics[width=3.3 in]{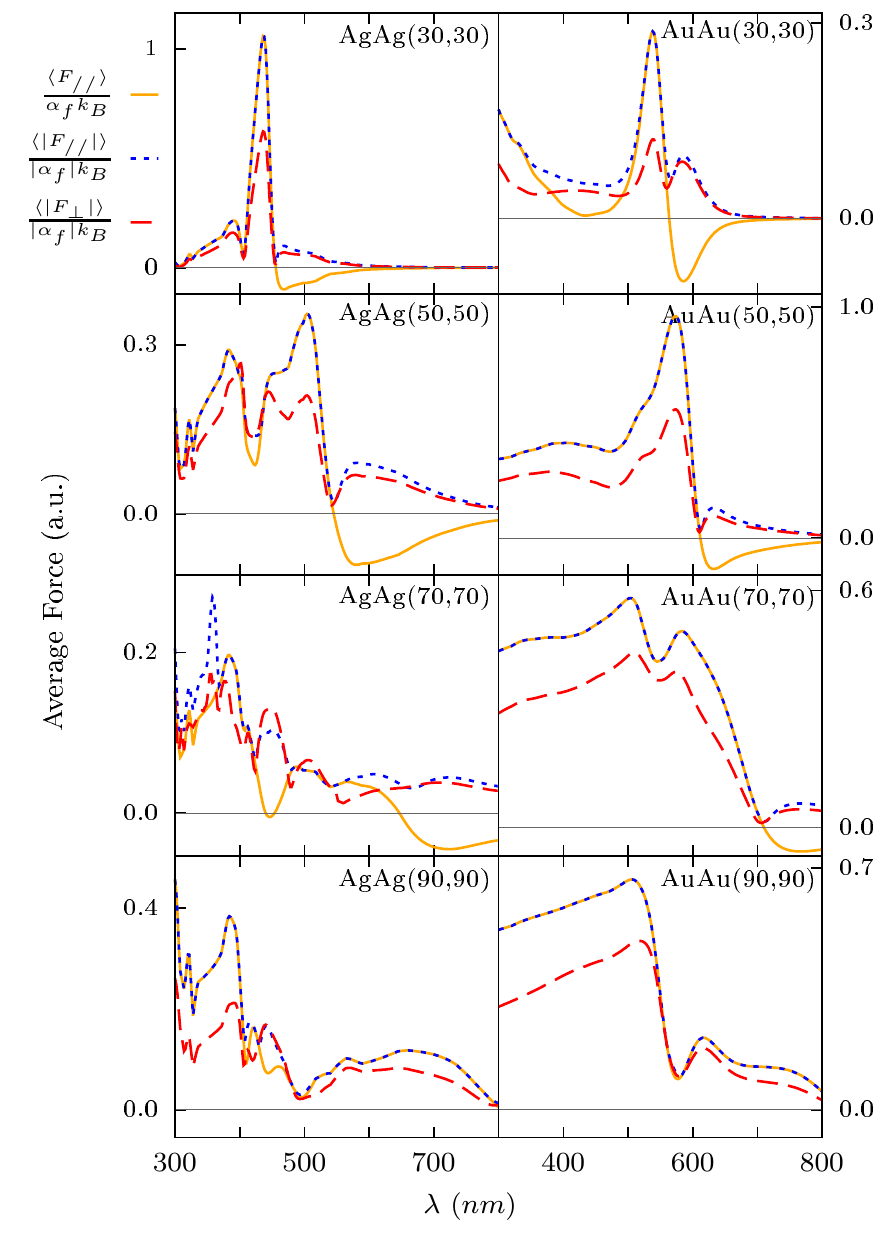} 
	\caption{Force spectra of Ag and Au homodimers with different radii.
	The gap is $5\,nm$ and we used the maximum of the force spectrum of AgAg(20,20) to normalize the figures.
	}
\label{fig:8}
\end{figure}

Fig. \ref{fig:8} shows the force spectra of symmetrical Ag and Au homodimers of different sizes.
In general, the Ag nanodimers with a larger radius show more complex force profiles than those of Au nanodimers.
Another general characteristic is that there is a redshift, for larger nanodimers, of the spectral region at which there is a change in the sign of the average force ($\langle F_{//} \rangle$).
Comparing the different force spectra, we found that Ag nanodimers with radii between $20-40\ nm$ maximize the average temperature difference (the maximum value of $\langle F_{//} \rangle/\alpha_f k_B$ is almost the same for these radii), while for Au nanodimers, the optimal radius is $50\ nm$.

\section{Conclusion\label{sec:conclusions}}

We have shown that simple metallic nanodimers can be excellent candidates for controllable active particles.
As such, the studied system possesses several advantages. Firs, the proposed structures are some of the simplest structures that can be made in a laboratory.
Second, the thermophoretic forces present phototaxis and a strong dependence of $\langle F_{//} \rangle $ with $\lambda$.
This may allow controlling externally the direction of the ensemble of active particles by simply changing the direction or the frequency of the incident light,
which may be important,e.g., for applications involving drug-carrying.
Indeed, changing $\lambda$ can even lead to the change of the sign of the average force.
Moreover, the strong dependency of $\langle F_{//} \rangle $ with $\lambda$ could be used, in principle, to separate a particular nanodimer geometry from a mixture of monomers or even nanodimers of different sizes and shapes.

It is interesting that often in the literature of active particles it is assumed that an inherent asymmetry of the nanostructures is a mandatory requirement for the candidate systems. On the contrary, this work suggests that the more promising phototactic nanodimers are the most symmetrical: the nanodimers made of identical nanoparticles. The necessary mirror symmetry breaking comes in this case from the direction of the incident light and not from the structure itself.

Although more studies are required, especially to assess the values of the $\alpha_f$ coefficients, our results suggest that the best candidates for phototactic thermophoretic nanoswimmers are the symmetrical Ag homodimers with radii between $20nm$ and $40nm$.
They should have, comparatively, the largest average forces $\langle F_{//} \rangle$ (or at least the largest average temperature differences) with the relatively smallest values of $\langle | F_{\perp} | \rangle$, which is associated with the lateral dispersion of an illuminated ensemble of nanodimers.
Additionally, the force spectra of these nanodimers show a narrow peak, which should help to target nanodimers of a given size to control them. The Au homodimer with a radius of $50nm$ seems also like a good candidate.
It shows almost the same value of the average temperature difference as that of AgAg(20,20), but the peak is much wider in this case.

Despite that discussed above, heterodimers or asymmetrical homodimers could still be useful for other purposes.
Interestingly, those systems present a controllable diffusion but with minimal net displacement in general.
In this regard, our results suggest that the effective diffusion coefficients can be tuned not only by a variation of the size and composition of the NPs, but also by a variation of the laser's intensity or wavelength. This may be useful,e.g., for applications involving the control of the self-assembly of complex nanomaterials.

As we mentioned, in this work we were not interested in calculating the exact value of thermophoretic forces but in describing general properties of these forces in illuminated metallic nanodimers.
However, we believe our results may be useful for guiding further experimental and theoretical studies.
In this respect, we hope that the simplicity of the geometry studied encourages further studies, especially from the experimental point of view.

\section{ACKNOWLEDGEMENTS}

Ra\'ul Bustos-Mar\'un acknowledges useful discussions with Lucas Barberis.
This work was supported by Consejo Nacional de Investigaciones Cient\'ificas y T\'ecnicas (CONICET), Argentina; and by Secretar\'ia de Ciencia y Tecnolog\'ia de la Universidad Nacional de C\'ordoba (SECYT-UNC), C\'ordoba, Argentina.

\appendix
\section{Estimation of thermophoretic forces and velocities.\label{sec:Appendix}}
The net thermophoretic force $\vec{F}$ produced by a temperature gradient $\nabla T$ can be computed by\cite{xu2017} 
\begin{equation}
\vec{F}=-C \nabla T. 
\end{equation}
The coefficient $C$ is given by $(9 \pi R \eta^2 k_a)/(\rho T k_p)$, where $R$ is the radius of the particle, $\eta$ is the fluid's viscosity, $k_a$ is the fluid's thermal conductivity, $k_p$ is the particle's thermal conductivity, and $\rho$ is the fluid's density. We will take $\nabla T$ equal to the temperature difference between the NPs divided by the center to center distance between them,
\begin{equation}
 \vec{F} = - \left ( \frac{C}{2R+d_{gap}} \right )  \Delta T \hat n. \label{eq:F_approx}
\end{equation}
where $d_{gap}$ is the length of the gap between the NPs. 
The friction coefficient can be calculated by assuming the system as an ellipsoid moving at random \cite{bromberg2011}
\begin{equation}
 \gamma = 6 \pi \eta \frac{a}{\ln (2 a/ b)}, \label{eq:g_sphe}
\end{equation}
where $a=4R+d_{gap}$ and $b=R$. 
Then, using $R=20 nm$, $d_{gap}=5 nm$, $k_a= 0.6 W m^{-1} K^{-1} $, $k_p= 427 W m^{-1} K^{-1}$, $T=293 K $, $\eta=10^{-3} Pa \cdot s $, $\rho = 10^3 Kg/m^3$, and $\Delta T = 7.3 \times 10^{-3} K$, we obtained $|F| = 0.44 \times 10^{-15} N$, $\gamma = 0.55 \times 10^{-9} Kg/s$, and $V = 0.8 \mu m/s$ ($\sim 10$ body lengths per second). For $V$ we use Eqs. \ref{eq:F_approx} and \ref{eq:g_sphe} on Eq. \ref{eq:V}, while for $\Delta T$ we take the value of $\left\langle F_{//}\right\rangle/\alpha_f k_B$ for the AgAg(20,20) nanodimer being illuminated by a laser with an irradiance of $10^{6}W/m^{2}$ at $\lambda=416nm$.

Of course, the above values of $F$ and $V$ are just crude estimates and deviations are expected. However, here we only wanted to emphasize that thermophoretic forces on metallic nanodimers can be important, especially considering that these systems present phototaxis and thus the velocities do not average to zero.
\bibliography{biblio-BPBM-arxiv-02}

\begin{thebibliography}{53}%
\makeatletter
\providecommand \@ifxundefined [1]{%
 \@ifx{#1\undefined}
}%
\providecommand \@ifnum [1]{%
 \ifnum #1\expandafter \@firstoftwo
 \else \expandafter \@secondoftwo
 \fi
}%
\providecommand \@ifx [1]{%
 \ifx #1\expandafter \@firstoftwo
 \else \expandafter \@secondoftwo
 \fi
}%
\providecommand \natexlab [1]{#1}%
\providecommand \enquote  [1]{``#1''}%
\providecommand \bibnamefont  [1]{#1}%
\providecommand \bibfnamefont [1]{#1}%
\providecommand \citenamefont [1]{#1}%
\providecommand \href@noop [0]{\@secondoftwo}%
\providecommand \href [0]{\begingroup \@sanitize@url \@href}%
\providecommand \@href[1]{\@@startlink{#1}\@@href}%
\providecommand \@@href[1]{\endgroup#1\@@endlink}%
\providecommand \@sanitize@url [0]{\catcode `\\12\catcode `\$12\catcode
  `\&12\catcode `\#12\catcode `\^12\catcode `\_12\catcode `\%12\relax}%
\providecommand \@@startlink[1]{}%
\providecommand \@@endlink[0]{}%
\providecommand \url  [0]{\begingroup\@sanitize@url \@url }%
\providecommand \@url [1]{\endgroup\@href {#1}{\urlprefix }}%
\providecommand \urlprefix  [0]{URL }%
\providecommand \Eprint [0]{\href }%
\providecommand \doibase [0]{http://dx.doi.org/}%
\providecommand \selectlanguage [0]{\@gobble}%
\providecommand \bibinfo  [0]{\@secondoftwo}%
\providecommand \bibfield  [0]{\@secondoftwo}%
\providecommand \translation [1]{[#1]}%
\providecommand \BibitemOpen [0]{}%
\providecommand \bibitemStop [0]{}%
\providecommand \bibitemNoStop [0]{.\EOS\space}%
\providecommand \EOS [0]{\spacefactor3000\relax}%
\providecommand \BibitemShut  [1]{\csname bibitem#1\endcsname}%
\let\auto@bib@innerbib\@empty
\bibitem [{\citenamefont {Golestanian}\ \emph {et~al.}(2007)\citenamefont
  {Golestanian}, \citenamefont {Liverpool},\ and\ \citenamefont
  {Ajdari}}]{golestanian2007}%
  \BibitemOpen
  \bibfield  {author} {\bibinfo {author} {\bibfnamefont {R.~.}\ \bibnamefont
  {Golestanian}}, \bibinfo {author} {\bibfnamefont {T.~B.}\ \bibnamefont
  {Liverpool}}, \ and\ \bibinfo {author} {\bibfnamefont {A.}~\bibnamefont
  {Ajdari}},\ }\href {\doibase 10.1088/1367-2630/9/5/126} {\bibfield  {journal}
  {\bibinfo  {journal} {New J. Phys.}\ }\textbf {\bibinfo {volume} {9}},\
  \bibinfo {pages} {126} (\bibinfo {year} {2007})}\BibitemShut {NoStop}%
\bibitem [{\citenamefont {Walther}\ and\ \citenamefont
  {M\"uller}(2008)}]{walther2008}%
  \BibitemOpen
  \bibfield  {author} {\bibinfo {author} {\bibfnamefont {A.}~\bibnamefont
  {Walther}}\ and\ \bibinfo {author} {\bibfnamefont {A.~H.~E.}\ \bibnamefont
  {M\"uller}},\ }\href {\doibase 10.1039/B718131K} {\bibfield  {journal}
  {\bibinfo  {journal} {Soft Matter}\ }\textbf {\bibinfo {volume} {4}},\
  \bibinfo {pages} {663} (\bibinfo {year} {2008})}\BibitemShut {NoStop}%
\bibitem [{\citenamefont {Wang}\ \emph {et~al.}(2013)\citenamefont {Wang},
  \citenamefont {Duan}, \citenamefont {Ahmed}, \citenamefont {Mallouk},\ and\
  \citenamefont {Sen}}]{wang2013}%
  \BibitemOpen
  \bibfield  {author} {\bibinfo {author} {\bibfnamefont {W.}~\bibnamefont
  {Wang}}, \bibinfo {author} {\bibfnamefont {W.}~\bibnamefont {Duan}}, \bibinfo
  {author} {\bibfnamefont {S.}~\bibnamefont {Ahmed}}, \bibinfo {author}
  {\bibfnamefont {T.~E.}\ \bibnamefont {Mallouk}}, \ and\ \bibinfo {author}
  {\bibfnamefont {A.}~\bibnamefont {Sen}},\ }\href {\doibase
  https://doi.org/10.1016/j.nantod.2013.08.009} {\bibfield  {journal} {\bibinfo
   {journal} {Nano Today}\ }\textbf {\bibinfo {volume} {8}},\ \bibinfo {pages}
  {531 } (\bibinfo {year} {2013})}\BibitemShut {NoStop}%
\bibitem [{\citenamefont {Wu}\ \emph {et~al.}(2016{\natexlab{a}})\citenamefont
  {Wu}, \citenamefont {Lin}, \citenamefont {Si},\ and\ \citenamefont
  {He}}]{wu2016rev}%
  \BibitemOpen
  \bibfield  {author} {\bibinfo {author} {\bibfnamefont {Z.}~\bibnamefont
  {Wu}}, \bibinfo {author} {\bibfnamefont {X.}~\bibnamefont {Lin}}, \bibinfo
  {author} {\bibfnamefont {T.}~\bibnamefont {Si}}, \ and\ \bibinfo {author}
  {\bibfnamefont {Q.}~\bibnamefont {He}},\ }\href {\doibase
  10.1002/smll.201503969} {\bibfield  {journal} {\bibinfo  {journal} {Small}\
  }\textbf {\bibinfo {volume} {12}},\ \bibinfo {pages} {3080} (\bibinfo {year}
  {2016}{\natexlab{a}})}\BibitemShut {NoStop}%
\bibitem [{\citenamefont {Lin}\ \emph {et~al.}(2017)\citenamefont {Lin},
  \citenamefont {Si}, \citenamefont {Wu},\ and\ \citenamefont {He}}]{lin2017}%
  \BibitemOpen
  \bibfield  {author} {\bibinfo {author} {\bibfnamefont {X.}~\bibnamefont
  {Lin}}, \bibinfo {author} {\bibfnamefont {T.}~\bibnamefont {Si}}, \bibinfo
  {author} {\bibfnamefont {Z.}~\bibnamefont {Wu}}, \ and\ \bibinfo {author}
  {\bibfnamefont {Q.}~\bibnamefont {He}},\ }\href {\doibase 10.1039/C7CP02561K}
  {\bibfield  {journal} {\bibinfo  {journal} {Phys. Chem. Chem. Phys.}\
  }\textbf {\bibinfo {volume} {19}},\ \bibinfo {pages} {23606} (\bibinfo {year}
  {2017})}\BibitemShut {NoStop}%
\bibitem [{\citenamefont {Moran}\ and\ \citenamefont
  {Posner}(2017)}]{moran2017}%
  \BibitemOpen
  \bibfield  {author} {\bibinfo {author} {\bibfnamefont {J.~L.}\ \bibnamefont
  {Moran}}\ and\ \bibinfo {author} {\bibfnamefont {J.~D.}\ \bibnamefont
  {Posner}},\ }\href {\doibase 10.1146/annurev-fluid-122414-034456} {\bibfield
  {journal} {\bibinfo  {journal} {Annu. Rev. Fluid Mech.}\ }\textbf {\bibinfo
  {volume} {49}},\ \bibinfo {pages} {511} (\bibinfo {year} {2017})}\BibitemShut
  {NoStop}%
\bibitem [{\citenamefont {Xu}\ \emph {et~al.}(2017)\citenamefont {Xu},
  \citenamefont {Mou}, \citenamefont {Gong}, \citenamefont {Luo},\ and\
  \citenamefont {Guan}}]{xu2017}%
  \BibitemOpen
  \bibfield  {author} {\bibinfo {author} {\bibfnamefont {L.}~\bibnamefont
  {Xu}}, \bibinfo {author} {\bibfnamefont {F.}~\bibnamefont {Mou}}, \bibinfo
  {author} {\bibfnamefont {H.}~\bibnamefont {Gong}}, \bibinfo {author}
  {\bibfnamefont {M.}~\bibnamefont {Luo}}, \ and\ \bibinfo {author}
  {\bibfnamefont {J.}~\bibnamefont {Guan}},\ }\href {\doibase
  10.1039/C7CS00516D} {\bibfield  {journal} {\bibinfo  {journal} {Chem. Soc.
  Rev.}\ }\textbf {\bibinfo {volume} {46}},\ \bibinfo {pages} {6905} (\bibinfo
  {year} {2017})}\BibitemShut {NoStop}%
\bibitem [{\citenamefont {Guix}\ \emph {et~al.}(2018)\citenamefont {Guix},
  \citenamefont {Weiz}, \citenamefont {Schmidt},\ and\ \citenamefont
  {Medina-S\'anchez}}]{guix2018}%
  \BibitemOpen
  \bibfield  {author} {\bibinfo {author} {\bibfnamefont {M.}~\bibnamefont
  {Guix}}, \bibinfo {author} {\bibfnamefont {S.~M.}\ \bibnamefont {Weiz}},
  \bibinfo {author} {\bibfnamefont {O.~G.}\ \bibnamefont {Schmidt}}, \ and\
  \bibinfo {author} {\bibfnamefont {M.}~\bibnamefont {Medina-S\'anchez}},\
  }\href {\doibase 10.1002/ppsc.201700382} {\bibfield  {journal} {\bibinfo
  {journal} {Part. Part. Syst. Charact.}\ }\textbf {\bibinfo {volume} {35}},\
  \bibinfo {pages} {1700382} (\bibinfo {year} {2018})}\BibitemShut {NoStop}%
\bibitem [{\citenamefont {Howse}\ \emph {et~al.}(2007)\citenamefont {Howse},
  \citenamefont {Jones}, \citenamefont {Ryan}, \citenamefont {Gough},
  \citenamefont {Vafabakhsh},\ and\ \citenamefont {Golestanian}}]{howse2007}%
  \BibitemOpen
  \bibfield  {author} {\bibinfo {author} {\bibfnamefont {J.~R.}\ \bibnamefont
  {Howse}}, \bibinfo {author} {\bibfnamefont {R.~A.~L.}\ \bibnamefont {Jones}},
  \bibinfo {author} {\bibfnamefont {A.~J.}\ \bibnamefont {Ryan}}, \bibinfo
  {author} {\bibfnamefont {T.}~\bibnamefont {Gough}}, \bibinfo {author}
  {\bibfnamefont {R.}~\bibnamefont {Vafabakhsh}}, \ and\ \bibinfo {author}
  {\bibfnamefont {R.}~\bibnamefont {Golestanian}},\ }\href {\doibase
  10.1103/PhysRevLett.99.048102} {\bibfield  {journal} {\bibinfo  {journal}
  {Phys. Rev. Lett.}\ }\textbf {\bibinfo {volume} {99}},\ \bibinfo {pages}
  {048102} (\bibinfo {year} {2007})}\BibitemShut {NoStop}%
\bibitem [{\citenamefont {R\"uckner}\ and\ \citenamefont
  {Kapral}(2007)}]{ruckner2007}%
  \BibitemOpen
  \bibfield  {author} {\bibinfo {author} {\bibfnamefont {G.}~\bibnamefont
  {R\"uckner}}\ and\ \bibinfo {author} {\bibfnamefont {R.}~\bibnamefont
  {Kapral}},\ }\href {\doibase 10.1103/PhysRevLett.98.150603} {\bibfield
  {journal} {\bibinfo  {journal} {Phys. Rev. Lett.}\ }\textbf {\bibinfo
  {volume} {98}},\ \bibinfo {pages} {150603} (\bibinfo {year}
  {2007})}\BibitemShut {NoStop}%
\bibitem [{\citenamefont {Lee}\ \emph {et~al.}(2014)\citenamefont {Lee},
  \citenamefont {Alarcón-Correa}, \citenamefont {Miksch}, \citenamefont
  {Hahn}, \citenamefont {Gibbs},\ and\ \citenamefont {Fischer}}]{lee2014}%
  \BibitemOpen
  \bibfield  {author} {\bibinfo {author} {\bibfnamefont {T.-C.}\ \bibnamefont
  {Lee}}, \bibinfo {author} {\bibfnamefont {M.}~\bibnamefont
  {Alarcón-Correa}}, \bibinfo {author} {\bibfnamefont {C.}~\bibnamefont
  {Miksch}}, \bibinfo {author} {\bibfnamefont {K.}~\bibnamefont {Hahn}},
  \bibinfo {author} {\bibfnamefont {J.~G.}\ \bibnamefont {Gibbs}}, \ and\
  \bibinfo {author} {\bibfnamefont {P.}~\bibnamefont {Fischer}},\ }\href
  {\doibase 10.1021/nl500068n} {\bibfield  {journal} {\bibinfo  {journal} {Nano
  Lett.}\ }\textbf {\bibinfo {volume} {14}},\ \bibinfo {pages} {2407} (\bibinfo
  {year} {2014})}\BibitemShut {NoStop}%
\bibitem [{\citenamefont {Schattling}\ \emph {et~al.}(2015)\citenamefont
  {Schattling}, \citenamefont {Thingholm},\ and\ \citenamefont
  {St\"adler}}]{schattling2015}%
  \BibitemOpen
  \bibfield  {author} {\bibinfo {author} {\bibfnamefont {P.}~\bibnamefont
  {Schattling}}, \bibinfo {author} {\bibfnamefont {B.}~\bibnamefont
  {Thingholm}}, \ and\ \bibinfo {author} {\bibfnamefont {B.}~\bibnamefont
  {St\"adler}},\ }\href {\doibase 10.1021/acs.chemmater.5b03303} {\bibfield
  {journal} {\bibinfo  {journal} {Chem. Mater.}\ }\textbf {\bibinfo {volume}
  {27}},\ \bibinfo {pages} {7412} (\bibinfo {year} {2015})}\BibitemShut
  {NoStop}%
\bibitem [{\citenamefont {Wang}\ \emph
  {et~al.}(2015{\natexlab{a}})\citenamefont {Wang}, \citenamefont {In},
  \citenamefont {Blanc}, \citenamefont {Nobili},\ and\ \citenamefont
  {Stocco}}]{wang2015}%
  \BibitemOpen
  \bibfield  {author} {\bibinfo {author} {\bibfnamefont {X.}~\bibnamefont
  {Wang}}, \bibinfo {author} {\bibfnamefont {M.}~\bibnamefont {In}}, \bibinfo
  {author} {\bibfnamefont {C.}~\bibnamefont {Blanc}}, \bibinfo {author}
  {\bibfnamefont {M.}~\bibnamefont {Nobili}}, \ and\ \bibinfo {author}
  {\bibfnamefont {A.}~\bibnamefont {Stocco}},\ }\href {\doibase
  10.1039/C5SM01111F} {\bibfield  {journal} {\bibinfo  {journal} {Soft Matter}\
  }\textbf {\bibinfo {volume} {11}},\ \bibinfo {pages} {7376} (\bibinfo {year}
  {2015}{\natexlab{a}})}\BibitemShut {NoStop}%
\bibitem [{\citenamefont {Qin}\ \emph {et~al.}(2017)\citenamefont {Qin},
  \citenamefont {Peng}, \citenamefont {Gao}, \citenamefont {Wang},
  \citenamefont {Hu}, \citenamefont {Wang}, \citenamefont {Shi}, \citenamefont
  {Li}, \citenamefont {Ren},\ and\ \citenamefont {Fan}}]{qin2017}%
  \BibitemOpen
  \bibfield  {author} {\bibinfo {author} {\bibfnamefont {W.}~\bibnamefont
  {Qin}}, \bibinfo {author} {\bibfnamefont {T.}~\bibnamefont {Peng}}, \bibinfo
  {author} {\bibfnamefont {Y.}~\bibnamefont {Gao}}, \bibinfo {author}
  {\bibfnamefont {F.}~\bibnamefont {Wang}}, \bibinfo {author} {\bibfnamefont
  {X.}~\bibnamefont {Hu}}, \bibinfo {author} {\bibfnamefont {K.}~\bibnamefont
  {Wang}}, \bibinfo {author} {\bibfnamefont {J.}~\bibnamefont {Shi}}, \bibinfo
  {author} {\bibfnamefont {D.}~\bibnamefont {Li}}, \bibinfo {author}
  {\bibfnamefont {J.}~\bibnamefont {Ren}}, \ and\ \bibinfo {author}
  {\bibfnamefont {C.}~\bibnamefont {Fan}},\ }\href {\doibase
  10.1002/anie.201609121} {\bibfield  {journal} {\bibinfo  {journal} {Angew.
  Chem. Int. Ed.}\ }\textbf {\bibinfo {volume} {56}},\ \bibinfo {pages} {515}
  (\bibinfo {year} {2017})}\BibitemShut {NoStop}%
\bibitem [{\citenamefont {Jiang}\ \emph {et~al.}(2010)\citenamefont {Jiang},
  \citenamefont {Yoshinaga},\ and\ \citenamefont {Sano}}]{jiang2010}%
  \BibitemOpen
  \bibfield  {author} {\bibinfo {author} {\bibfnamefont {H.-R.}\ \bibnamefont
  {Jiang}}, \bibinfo {author} {\bibfnamefont {N.}~\bibnamefont {Yoshinaga}}, \
  and\ \bibinfo {author} {\bibfnamefont {M.}~\bibnamefont {Sano}},\ }\href
  {\doibase 10.1103/PhysRevLett.105.268302} {\bibfield  {journal} {\bibinfo
  {journal} {Phys. Rev. Lett.}\ }\textbf {\bibinfo {volume} {105}},\ \bibinfo
  {pages} {268302} (\bibinfo {year} {2010})}\BibitemShut {NoStop}%
\bibitem [{\citenamefont {Buttinoni}\ \emph {et~al.}(2013)\citenamefont
  {Buttinoni}, \citenamefont {Bialk\'e}, \citenamefont {K\"ummel},
  \citenamefont {L\"owen}, \citenamefont {Bechinger},\ and\ \citenamefont
  {Speck}}]{buttinoni2013}%
  \BibitemOpen
  \bibfield  {author} {\bibinfo {author} {\bibfnamefont {I.}~\bibnamefont
  {Buttinoni}}, \bibinfo {author} {\bibfnamefont {J.}~\bibnamefont {Bialk\'e}},
  \bibinfo {author} {\bibfnamefont {F.}~\bibnamefont {K\"ummel}}, \bibinfo
  {author} {\bibfnamefont {H.}~\bibnamefont {L\"owen}}, \bibinfo {author}
  {\bibfnamefont {C.}~\bibnamefont {Bechinger}}, \ and\ \bibinfo {author}
  {\bibfnamefont {T.}~\bibnamefont {Speck}},\ }\href {\doibase
  10.1103/PhysRevLett.110.238301} {\bibfield  {journal} {\bibinfo  {journal}
  {Phys. Rev. Lett.}\ }\textbf {\bibinfo {volume} {110}},\ \bibinfo {pages}
  {238301} (\bibinfo {year} {2013})}\BibitemShut {NoStop}%
\bibitem [{\citenamefont {Baraban}\ \emph {et~al.}(2013)\citenamefont
  {Baraban}, \citenamefont {Streubel}, \citenamefont {Makarov}, \citenamefont
  {Han}, \citenamefont {Karnaushenko}, \citenamefont {Schmidt},\ and\
  \citenamefont {Cuniberti}}]{baraban2013}%
  \BibitemOpen
  \bibfield  {author} {\bibinfo {author} {\bibfnamefont {L.}~\bibnamefont
  {Baraban}}, \bibinfo {author} {\bibfnamefont {R.}~\bibnamefont {Streubel}},
  \bibinfo {author} {\bibfnamefont {D.}~\bibnamefont {Makarov}}, \bibinfo
  {author} {\bibfnamefont {L.}~\bibnamefont {Han}}, \bibinfo {author}
  {\bibfnamefont {D.}~\bibnamefont {Karnaushenko}}, \bibinfo {author}
  {\bibfnamefont {O.~G.}\ \bibnamefont {Schmidt}}, \ and\ \bibinfo {author}
  {\bibfnamefont {G.}~\bibnamefont {Cuniberti}},\ }\href {\doibase
  10.1021/nn305726m} {\bibfield  {journal} {\bibinfo  {journal} {ACS Nano}\
  }\textbf {\bibinfo {volume} {7}},\ \bibinfo {pages} {1360} (\bibinfo {year}
  {2013})}\BibitemShut {NoStop}%
\bibitem [{\citenamefont {K\"ummel}\ \emph {et~al.}(2013)\citenamefont
  {K\"ummel}, \citenamefont {ten Hagen}, \citenamefont {Wittkowski},
  \citenamefont {Buttinoni}, \citenamefont {Eichhorn}, \citenamefont {Volpe},
  \citenamefont {L\"owen},\ and\ \citenamefont {Bechinger}}]{kummel2013}%
  \BibitemOpen
  \bibfield  {author} {\bibinfo {author} {\bibfnamefont {F.}~\bibnamefont
  {K\"ummel}}, \bibinfo {author} {\bibfnamefont {B.}~\bibnamefont {ten Hagen}},
  \bibinfo {author} {\bibfnamefont {R.}~\bibnamefont {Wittkowski}}, \bibinfo
  {author} {\bibfnamefont {I.}~\bibnamefont {Buttinoni}}, \bibinfo {author}
  {\bibfnamefont {R.}~\bibnamefont {Eichhorn}}, \bibinfo {author}
  {\bibfnamefont {G.}~\bibnamefont {Volpe}}, \bibinfo {author} {\bibfnamefont
  {H.}~\bibnamefont {L\"owen}}, \ and\ \bibinfo {author} {\bibfnamefont
  {C.}~\bibnamefont {Bechinger}},\ }\href {\doibase
  10.1103/PhysRevLett.110.198302} {\bibfield  {journal} {\bibinfo  {journal}
  {Phys. Rev. Lett.}\ }\textbf {\bibinfo {volume} {110}},\ \bibinfo {pages}
  {198302} (\bibinfo {year} {2013})}\BibitemShut {NoStop}%
\bibitem [{\citenamefont {Wang}\ \emph
  {et~al.}(2015{\natexlab{b}})\citenamefont {Wang}, \citenamefont {Duan},
  \citenamefont {Ahmed}, \citenamefont {Sen},\ and\ \citenamefont
  {Mallouk}}]{wang2015b}%
  \BibitemOpen
  \bibfield  {author} {\bibinfo {author} {\bibfnamefont {W.}~\bibnamefont
  {Wang}}, \bibinfo {author} {\bibfnamefont {W.}~\bibnamefont {Duan}}, \bibinfo
  {author} {\bibfnamefont {S.}~\bibnamefont {Ahmed}}, \bibinfo {author}
  {\bibfnamefont {A.}~\bibnamefont {Sen}}, \ and\ \bibinfo {author}
  {\bibfnamefont {T.~E.}\ \bibnamefont {Mallouk}},\ }\href {\doibase
  10.1021/acs.accounts.5b00025} {\bibfield  {journal} {\bibinfo  {journal}
  {Acc. Chem. Res.}\ }\textbf {\bibinfo {volume} {48}},\ \bibinfo {pages}
  {1938} (\bibinfo {year} {2015}{\natexlab{b}})}\BibitemShut {NoStop}%
\bibitem [{\citenamefont {Golestanian}(2012)}]{golestanian2012}%
  \BibitemOpen
  \bibfield  {author} {\bibinfo {author} {\bibfnamefont {R.}~\bibnamefont
  {Golestanian}},\ }\href@noop {} {\bibfield  {journal} {\bibinfo  {journal}
  {Phys. Rev. Lett.}\ }\textbf {\bibinfo {volume} {108}},\ \bibinfo {pages}
  {038303} (\bibinfo {year} {2012})}\BibitemShut {NoStop}%
\bibitem [{\citenamefont {Metwally}\ \emph {et~al.}(2015)\citenamefont
  {Metwally}, \citenamefont {Mensah},\ and\ \citenamefont
  {Baffou}}]{metwally2015}%
  \BibitemOpen
  \bibfield  {author} {\bibinfo {author} {\bibfnamefont {K.}~\bibnamefont
  {Metwally}}, \bibinfo {author} {\bibfnamefont {S.}~\bibnamefont {Mensah}}, \
  and\ \bibinfo {author} {\bibfnamefont {G.}~\bibnamefont {Baffou}},\ }\href
  {\doibase 10.1021/acs.jpcc.5b09903} {\bibfield  {journal} {\bibinfo
  {journal} {J. Phys. Chem. C}\ }\textbf {\bibinfo {volume} {119}},\ \bibinfo
  {pages} {28586} (\bibinfo {year} {2015})}\BibitemShut {NoStop}%
\bibitem [{\citenamefont {Frenkel}\ and\ \citenamefont
  {Niv}(2017)}]{frenkel2017}%
  \BibitemOpen
  \bibfield  {author} {\bibinfo {author} {\bibfnamefont {I.}~\bibnamefont
  {Frenkel}}\ and\ \bibinfo {author} {\bibfnamefont {A.}~\bibnamefont {Niv}},\
  }\href@noop {} {\bibfield  {journal} {\bibinfo  {journal} {Sci. Rep.}\
  }\textbf {\bibinfo {volume} {7}},\ \bibinfo {pages} {2814} (\bibinfo {year}
  {2017})}\BibitemShut {NoStop}%
\bibitem [{\citenamefont {Buttinoni}\ \emph {et~al.}(2012)\citenamefont
  {Buttinoni}, \citenamefont {Volpe}, \citenamefont {K{\"u}mmel}, \citenamefont
  {Volpe},\ and\ \citenamefont {Bechinger}}]{buttinoni2012}%
  \BibitemOpen
  \bibfield  {author} {\bibinfo {author} {\bibfnamefont {I.}~\bibnamefont
  {Buttinoni}}, \bibinfo {author} {\bibfnamefont {G.}~\bibnamefont {Volpe}},
  \bibinfo {author} {\bibfnamefont {F.}~\bibnamefont {K{\"u}mmel}}, \bibinfo
  {author} {\bibfnamefont {G.}~\bibnamefont {Volpe}}, \ and\ \bibinfo {author}
  {\bibfnamefont {C.}~\bibnamefont {Bechinger}},\ }\href@noop {} {\bibfield
  {journal} {\bibinfo  {journal} {J. Phys. Condens. Matter}\ }\textbf {\bibinfo
  {volume} {24}},\ \bibinfo {pages} {284129} (\bibinfo {year}
  {2012})}\BibitemShut {NoStop}%
\bibitem [{\citenamefont {Ma}\ \emph {et~al.}(2016)\citenamefont {Ma},
  \citenamefont {Jang}, \citenamefont {Popescu}, \citenamefont {Uspal},
  \citenamefont {Miguel-L\'opez}, \citenamefont {Hahn}, \citenamefont {Kim},\
  and\ \citenamefont {S\'anchez}}]{ma2016}%
  \BibitemOpen
  \bibfield  {author} {\bibinfo {author} {\bibfnamefont {X.}~\bibnamefont
  {Ma}}, \bibinfo {author} {\bibfnamefont {S.}~\bibnamefont {Jang}}, \bibinfo
  {author} {\bibfnamefont {M.~N.}\ \bibnamefont {Popescu}}, \bibinfo {author}
  {\bibfnamefont {W.~E.}\ \bibnamefont {Uspal}}, \bibinfo {author}
  {\bibfnamefont {A.}~\bibnamefont {Miguel-L\'opez}}, \bibinfo {author}
  {\bibfnamefont {K.}~\bibnamefont {Hahn}}, \bibinfo {author} {\bibfnamefont
  {D.-P.}\ \bibnamefont {Kim}}, \ and\ \bibinfo {author} {\bibfnamefont
  {S.}~\bibnamefont {S\'anchez}},\ }\href@noop {} {\bibfield  {journal}
  {\bibinfo  {journal} {ACS nano}\ }\textbf {\bibinfo {volume} {10}},\ \bibinfo
  {pages} {8751} (\bibinfo {year} {2016})}\BibitemShut {NoStop}%
\bibitem [{\citenamefont {Yang}\ and\ \citenamefont {Ripoll}(2011)}]{yang2011}%
  \BibitemOpen
  \bibfield  {author} {\bibinfo {author} {\bibfnamefont {M.}~\bibnamefont
  {Yang}}\ and\ \bibinfo {author} {\bibfnamefont {M.}~\bibnamefont {Ripoll}},\
  }\href@noop {} {\bibfield  {journal} {\bibinfo  {journal} {Phys. Rev. E}\
  }\textbf {\bibinfo {volume} {84}},\ \bibinfo {pages} {061401} (\bibinfo
  {year} {2011})}\BibitemShut {NoStop}%
\bibitem [{\citenamefont {Yang}\ \emph {et~al.}(2014)\citenamefont {Yang},
  \citenamefont {Wysocki},\ and\ \citenamefont {Ripoll}}]{yang2014}%
  \BibitemOpen
  \bibfield  {author} {\bibinfo {author} {\bibfnamefont {M.}~\bibnamefont
  {Yang}}, \bibinfo {author} {\bibfnamefont {A.}~\bibnamefont {Wysocki}}, \
  and\ \bibinfo {author} {\bibfnamefont {M.}~\bibnamefont {Ripoll}},\
  }\href@noop {} {\bibfield  {journal} {\bibinfo  {journal} {Soft matter}\
  }\textbf {\bibinfo {volume} {10}},\ \bibinfo {pages} {6208} (\bibinfo {year}
  {2014})}\BibitemShut {NoStop}%
\bibitem [{\citenamefont {Michaelides}(2015)}]{michaelides2015}%
  \BibitemOpen
  \bibfield  {author} {\bibinfo {author} {\bibfnamefont {E.~E.}\ \bibnamefont
  {Michaelides}},\ }\href {\doibase 10.1016/j.ijheatmasstransfer.2014.10.019}
  {\bibfield  {journal} {\bibinfo  {journal} {Int. J. Heat Mass Transf.}\
  }\textbf {\bibinfo {volume} {81}},\ \bibinfo {pages} {179 } (\bibinfo {year}
  {2015})}\BibitemShut {NoStop}%
\bibitem [{\citenamefont {Ilic}\ \emph {et~al.}(2016)\citenamefont {Ilic},
  \citenamefont {Kaminer}, \citenamefont {Lahini}, \citenamefont {Buljan},\
  and\ \citenamefont {Solja\u{c}i\'{c}}}]{ilic2016}%
  \BibitemOpen
  \bibfield  {author} {\bibinfo {author} {\bibfnamefont {O.}~\bibnamefont
  {Ilic}}, \bibinfo {author} {\bibfnamefont {I.}~\bibnamefont {Kaminer}},
  \bibinfo {author} {\bibfnamefont {Y.}~\bibnamefont {Lahini}}, \bibinfo
  {author} {\bibfnamefont {H.}~\bibnamefont {Buljan}}, \ and\ \bibinfo {author}
  {\bibfnamefont {M.}~\bibnamefont {Solja\u{c}i\'{c}}},\ }\href {\doibase
  10.1021/acsphotonics.5b00605} {\bibfield  {journal} {\bibinfo  {journal} {ACS
  Photonics}\ }\textbf {\bibinfo {volume} {3}},\ \bibinfo {pages} {197}
  (\bibinfo {year} {2016})}\BibitemShut {NoStop}%
\bibitem [{\citenamefont {Wu}\ \emph {et~al.}(2016{\natexlab{b}})\citenamefont
  {Wu}, \citenamefont {Si}, \citenamefont {Gao}, \citenamefont {Lin},
  \citenamefont {Wang},\ and\ \citenamefont {He}}]{wu2016}%
  \BibitemOpen
  \bibfield  {author} {\bibinfo {author} {\bibfnamefont {Z.}~\bibnamefont
  {Wu}}, \bibinfo {author} {\bibfnamefont {T.}~\bibnamefont {Si}}, \bibinfo
  {author} {\bibfnamefont {W.}~\bibnamefont {Gao}}, \bibinfo {author}
  {\bibfnamefont {X.}~\bibnamefont {Lin}}, \bibinfo {author} {\bibfnamefont
  {J.}~\bibnamefont {Wang}}, \ and\ \bibinfo {author} {\bibfnamefont
  {Q.}~\bibnamefont {He}},\ }\href {\doibase 10.1002/smll.201502605} {\bibfield
   {journal} {\bibinfo  {journal} {Small}\ }\textbf {\bibinfo {volume} {12}},\
  \bibinfo {pages} {577} (\bibinfo {year} {2016}{\natexlab{b}})}\BibitemShut
  {NoStop}%
\bibitem [{\citenamefont {Xuan}\ \emph {et~al.}(2016)\citenamefont {Xuan},
  \citenamefont {Wu}, \citenamefont {Shao}, \citenamefont {Dai}, \citenamefont
  {Si},\ and\ \citenamefont {He}}]{xuan2016}%
  \BibitemOpen
  \bibfield  {author} {\bibinfo {author} {\bibfnamefont {M.}~\bibnamefont
  {Xuan}}, \bibinfo {author} {\bibfnamefont {Z.}~\bibnamefont {Wu}}, \bibinfo
  {author} {\bibfnamefont {J.}~\bibnamefont {Shao}}, \bibinfo {author}
  {\bibfnamefont {L.}~\bibnamefont {Dai}}, \bibinfo {author} {\bibfnamefont
  {T.}~\bibnamefont {Si}}, \ and\ \bibinfo {author} {\bibfnamefont
  {Q.}~\bibnamefont {He}},\ }\href {\doibase 10.1021/jacs.6b00902} {\bibfield
  {journal} {\bibinfo  {journal} {J. Am. Chem. Soc.}\ }\textbf {\bibinfo
  {volume} {138}},\ \bibinfo {pages} {6492} (\bibinfo {year}
  {2016})}\BibitemShut {NoStop}%
\bibitem [{\citenamefont {Lan}\ \emph {et~al.}(2013)\citenamefont {Lan},
  \citenamefont {Chen}, \citenamefont {Liu}, \citenamefont {Ren}, \citenamefont
  {Henzie},\ and\ \citenamefont {Wang}}]{lan2013}%
  \BibitemOpen
  \bibfield  {author} {\bibinfo {author} {\bibfnamefont {X.}~\bibnamefont
  {Lan}}, \bibinfo {author} {\bibfnamefont {Z.}~\bibnamefont {Chen}}, \bibinfo
  {author} {\bibfnamefont {B.-J.}\ \bibnamefont {Liu}}, \bibinfo {author}
  {\bibfnamefont {B.}~\bibnamefont {Ren}}, \bibinfo {author} {\bibfnamefont
  {J.}~\bibnamefont {Henzie}}, \ and\ \bibinfo {author} {\bibfnamefont
  {Q.}~\bibnamefont {Wang}},\ }\href@noop {} {\bibfield  {journal} {\bibinfo
  {journal} {Small}\ }\textbf {\bibinfo {volume} {9}},\ \bibinfo {pages} {2308}
  (\bibinfo {year} {2013})}\BibitemShut {NoStop}%
\bibitem [{\citenamefont {Sheikholeslami}\ \emph {et~al.}(2010)\citenamefont
  {Sheikholeslami}, \citenamefont {Jun}, \citenamefont {Jain},\ and\
  \citenamefont {Alivisatos}}]{sheikholeslami2010}%
  \BibitemOpen
  \bibfield  {author} {\bibinfo {author} {\bibfnamefont {S.}~\bibnamefont
  {Sheikholeslami}}, \bibinfo {author} {\bibfnamefont {Y.-w.}\ \bibnamefont
  {Jun}}, \bibinfo {author} {\bibfnamefont {P.~K.}\ \bibnamefont {Jain}}, \
  and\ \bibinfo {author} {\bibfnamefont {A.~P.}\ \bibnamefont {Alivisatos}},\
  }\href@noop {} {\bibfield  {journal} {\bibinfo  {journal} {Nano lett.}\
  }\textbf {\bibinfo {volume} {10}},\ \bibinfo {pages} {2655} (\bibinfo {year}
  {2010})}\BibitemShut {NoStop}%
\bibitem [{\citenamefont {Sun}(2015)}]{sun2015}%
  \BibitemOpen
  \bibfield  {author} {\bibinfo {author} {\bibfnamefont {Y.}~\bibnamefont
  {Sun}},\ }\href {\doibase 10.1093/nsr/nwv037} {\bibfield  {journal} {\bibinfo
   {journal} {Natl. Sci. rev.}\ }\textbf {\bibinfo {volume} {2}},\ \bibinfo
  {pages} {329} (\bibinfo {year} {2015})}\BibitemShut {NoStop}%
\bibitem [{\citenamefont {Zheng}\ \emph {et~al.}(2015)\citenamefont {Zheng},
  \citenamefont {Rosa}, \citenamefont {Thai}, \citenamefont {Ng}, \citenamefont
  {G{\'o}mez}, \citenamefont {Ohshima},\ and\ \citenamefont
  {Bach}}]{zheng2015}%
  \BibitemOpen
  \bibfield  {author} {\bibinfo {author} {\bibfnamefont {Y.}~\bibnamefont
  {Zheng}}, \bibinfo {author} {\bibfnamefont {L.}~\bibnamefont {Rosa}},
  \bibinfo {author} {\bibfnamefont {T.}~\bibnamefont {Thai}}, \bibinfo {author}
  {\bibfnamefont {S.~H.}\ \bibnamefont {Ng}}, \bibinfo {author} {\bibfnamefont
  {D.~E.}\ \bibnamefont {G{\'o}mez}}, \bibinfo {author} {\bibfnamefont
  {H.}~\bibnamefont {Ohshima}}, \ and\ \bibinfo {author} {\bibfnamefont
  {U.}~\bibnamefont {Bach}},\ }\href@noop {} {\bibfield  {journal} {\bibinfo
  {journal} {J. Mater. Chem. A}\ }\textbf {\bibinfo {volume} {3}},\ \bibinfo
  {pages} {240} (\bibinfo {year} {2015})}\BibitemShut {NoStop}%
\bibitem [{\citenamefont {Jiang}\ \emph {et~al.}(2019)\citenamefont {Jiang},
  \citenamefont {Ji}, \citenamefont {Riley},\ and\ \citenamefont
  {Xie}}]{jiang2019}%
  \BibitemOpen
  \bibfield  {author} {\bibinfo {author} {\bibfnamefont {Q.}~\bibnamefont
  {Jiang}}, \bibinfo {author} {\bibfnamefont {C.}~\bibnamefont {Ji}}, \bibinfo
  {author} {\bibfnamefont {D.}~\bibnamefont {Riley}}, \ and\ \bibinfo {author}
  {\bibfnamefont {F.}~\bibnamefont {Xie}},\ }\href@noop {} {\bibfield
  {journal} {\bibinfo  {journal} {Nanomaterials}\ }\textbf {\bibinfo {volume}
  {9}},\ \bibinfo {pages} {1} (\bibinfo {year} {2019})}\BibitemShut {NoStop}%
\bibitem [{\citenamefont {Garc{\'\i}a-Negrete}\ \emph
  {et~al.}(2014)\citenamefont {Garc{\'\i}a-Negrete}, \citenamefont {Rojas},
  \citenamefont {Knappett}, \citenamefont {Jefferson}, \citenamefont
  {Wheatley},\ and\ \citenamefont {Fern{\'a}ndez}}]{garcia2014}%
  \BibitemOpen
  \bibfield  {author} {\bibinfo {author} {\bibfnamefont {C.~A.}\ \bibnamefont
  {Garc{\'\i}a-Negrete}}, \bibinfo {author} {\bibfnamefont {T.~C.}\
  \bibnamefont {Rojas}}, \bibinfo {author} {\bibfnamefont {B.~R.}\ \bibnamefont
  {Knappett}}, \bibinfo {author} {\bibfnamefont {D.~A.}\ \bibnamefont
  {Jefferson}}, \bibinfo {author} {\bibfnamefont {A.~E.}\ \bibnamefont
  {Wheatley}}, \ and\ \bibinfo {author} {\bibfnamefont {A.}~\bibnamefont
  {Fern{\'a}ndez}},\ }\href@noop {} {\bibfield  {journal} {\bibinfo  {journal}
  {Nanoscale}\ }\textbf {\bibinfo {volume} {6}},\ \bibinfo {pages} {11090}
  (\bibinfo {year} {2014})}\BibitemShut {NoStop}%
\bibitem [{\citenamefont {Baffou}\ \emph
  {et~al.}(2010{\natexlab{a}})\citenamefont {Baffou}, \citenamefont {Quidant},\
  and\ \citenamefont {Girard}}]{baffou2010}%
  \BibitemOpen
  \bibfield  {author} {\bibinfo {author} {\bibfnamefont {G.}~\bibnamefont
  {Baffou}}, \bibinfo {author} {\bibfnamefont {R.}~\bibnamefont {Quidant}}, \
  and\ \bibinfo {author} {\bibfnamefont {C.}~\bibnamefont {Girard}},\
  }\href@noop {} {\bibfield  {journal} {\bibinfo  {journal} {Phy. Rev. B}\
  }\textbf {\bibinfo {volume} {82}},\ \bibinfo {pages} {165424} (\bibinfo
  {year} {2010}{\natexlab{a}})}\BibitemShut {NoStop}%
\bibitem [{\citenamefont {Sabass}\ and\ \citenamefont
  {Seifert}(2010)}]{sabass2010}%
  \BibitemOpen
  \bibfield  {author} {\bibinfo {author} {\bibfnamefont {B.}~\bibnamefont
  {Sabass}}\ and\ \bibinfo {author} {\bibfnamefont {U.}~\bibnamefont
  {Seifert}},\ }\href@noop {} {\bibfield  {journal} {\bibinfo  {journal} {Phys.
  Rev. Lett.}\ }\textbf {\bibinfo {volume} {105}},\ \bibinfo {pages} {218103}
  (\bibinfo {year} {2010})}\BibitemShut {NoStop}%
\bibitem [{\citenamefont {Xu}(1997)}]{xu1997}%
  \BibitemOpen
  \bibfield  {author} {\bibinfo {author} {\bibfnamefont {Y.-l.}\ \bibnamefont
  {Xu}},\ }\href@noop {} {\bibfield  {journal} {\bibinfo  {journal} {Applied
  optics}\ }\textbf {\bibinfo {volume} {36}},\ \bibinfo {pages} {9496}
  (\bibinfo {year} {1997})}\BibitemShut {NoStop}%
\bibitem [{\citenamefont {Pellegrini}\ \emph {et~al.}(2007)\citenamefont
  {Pellegrini}, \citenamefont {Mattei}, \citenamefont {Bello},\ and\
  \citenamefont {Mazzoldi}}]{pellegrini2007}%
  \BibitemOpen
  \bibfield  {author} {\bibinfo {author} {\bibfnamefont {G.}~\bibnamefont
  {Pellegrini}}, \bibinfo {author} {\bibfnamefont {G.}~\bibnamefont {Mattei}},
  \bibinfo {author} {\bibfnamefont {V.}~\bibnamefont {Bello}}, \ and\ \bibinfo
  {author} {\bibfnamefont {P.}~\bibnamefont {Mazzoldi}},\ }\href@noop {}
  {\bibfield  {journal} {\bibinfo  {journal} {Mater. Sci. Eng. C}\ }\textbf
  {\bibinfo {volume} {27}},\ \bibinfo {pages} {1347} (\bibinfo {year}
  {2007})}\BibitemShut {NoStop}%
\bibitem [{\citenamefont {Bohren}(2008)}]{bohren2008}%
  \BibitemOpen
  \bibfield  {author} {\bibinfo {author} {\bibfnamefont {D.~R.}\ \bibnamefont
  {Bohren}, \bibfnamefont {C.~F.and~Huffman}},\ }\href@noop {} {\emph {\bibinfo
  {title} {Absorption and scattering of light by small particles}}}\ (\bibinfo
  {publisher} {Wiley Interscience, New York, USA},\ \bibinfo {year}
  {2008})\BibitemShut {NoStop}%
\bibitem [{\citenamefont {Bechinger}\ \emph {et~al.}(2016)\citenamefont
  {Bechinger}, \citenamefont {Di~Leonardo}, \citenamefont {L\"owen},
  \citenamefont {Reichhardt}, \citenamefont {Volpe},\ and\ \citenamefont
  {Volpe}}]{bechinger2016}%
  \BibitemOpen
  \bibfield  {author} {\bibinfo {author} {\bibfnamefont {C.}~\bibnamefont
  {Bechinger}}, \bibinfo {author} {\bibfnamefont {R.}~\bibnamefont
  {Di~Leonardo}}, \bibinfo {author} {\bibfnamefont {H.}~\bibnamefont
  {L\"owen}}, \bibinfo {author} {\bibfnamefont {C.}~\bibnamefont {Reichhardt}},
  \bibinfo {author} {\bibfnamefont {G.}~\bibnamefont {Volpe}}, \ and\ \bibinfo
  {author} {\bibfnamefont {G.}~\bibnamefont {Volpe}},\ }\href {\doibase
  10.1103/RevModPhys.88.045006} {\bibfield  {journal} {\bibinfo  {journal}
  {Rev. Mod. Phys.}\ }\textbf {\bibinfo {volume} {88}},\ \bibinfo {pages}
  {045006} (\bibinfo {year} {2016})}\BibitemShut {NoStop}%
\bibitem [{\citenamefont {Hale}\ and\ \citenamefont {Querry}(1973)}]{hale1973}%
  \BibitemOpen
  \bibfield  {author} {\bibinfo {author} {\bibfnamefont {G.~M.}\ \bibnamefont
  {Hale}}\ and\ \bibinfo {author} {\bibfnamefont {M.~R.}\ \bibnamefont
  {Querry}},\ }\href@noop {} {\bibfield  {journal} {\bibinfo  {journal} {Appl.
  Opt.}\ }\textbf {\bibinfo {volume} {12}},\ \bibinfo {pages} {555} (\bibinfo
  {year} {1973})}\BibitemShut {NoStop}%
\bibitem [{\citenamefont {Ferreiro-Vila}\ \emph {et~al.}(2009)\citenamefont
  {Ferreiro-Vila}, \citenamefont {Gonz\'alez-D\'iaz}, \citenamefont {Fermento},
  \citenamefont {Gonz\'alez}, \citenamefont {Garc\'ia-Mart\'in}, \citenamefont
  {Garc\'ia-Mart\'in}, \citenamefont {Cebollada}, \citenamefont {Armelles},
  \citenamefont {Meneses-Rodr\'iguez},\ and\ \citenamefont
  {Sandoval}}]{ferreiro2009}%
  \BibitemOpen
  \bibfield  {author} {\bibinfo {author} {\bibfnamefont {E.}~\bibnamefont
  {Ferreiro-Vila}}, \bibinfo {author} {\bibfnamefont {J.~B.}\ \bibnamefont
  {Gonz\'alez-D\'iaz}}, \bibinfo {author} {\bibfnamefont {R.}~\bibnamefont
  {Fermento}}, \bibinfo {author} {\bibfnamefont {M.~U.}\ \bibnamefont
  {Gonz\'alez}}, \bibinfo {author} {\bibfnamefont {A.}~\bibnamefont
  {Garc\'ia-Mart\'in}}, \bibinfo {author} {\bibfnamefont {J.~M.}\ \bibnamefont
  {Garc\'ia-Mart\'in}}, \bibinfo {author} {\bibfnamefont {A.}~\bibnamefont
  {Cebollada}}, \bibinfo {author} {\bibfnamefont {G.}~\bibnamefont {Armelles}},
  \bibinfo {author} {\bibfnamefont {D.}~\bibnamefont {Meneses-Rodr\'iguez}}, \
  and\ \bibinfo {author} {\bibfnamefont {E.~M.}\ \bibnamefont {Sandoval}},\
  }\href@noop {} {\bibfield  {journal} {\bibinfo  {journal} {Phys. Rev. B}\
  }\textbf {\bibinfo {volume} {80}},\ \bibinfo {pages} {125132} (\bibinfo
  {year} {2009})}\BibitemShut {NoStop}%
\bibitem [{\citenamefont {Johnson}\ and\ \citenamefont
  {Christy}(1972)}]{johnson1972}%
  \BibitemOpen
  \bibfield  {author} {\bibinfo {author} {\bibfnamefont {P.~B.}\ \bibnamefont
  {Johnson}}\ and\ \bibinfo {author} {\bibfnamefont {R.-W.}\ \bibnamefont
  {Christy}},\ }\href@noop {} {\bibfield  {journal} {\bibinfo  {journal} {Phys.
  Rev B}\ }\textbf {\bibinfo {volume} {6}},\ \bibinfo {pages} {4370} (\bibinfo
  {year} {1972})}\BibitemShut {NoStop}%
\bibitem [{\citenamefont {Palik}(1998)}]{palik1998}%
  \BibitemOpen
  \bibfield  {author} {\bibinfo {author} {\bibfnamefont {E.~D.}\ \bibnamefont
  {Palik}},\ }\href@noop {} {\emph {\bibinfo {title} {Handbook of Optical
  Constants of Solids}}}\ (\bibinfo  {publisher} {Academic Press},\ \bibinfo
  {year} {1998})\BibitemShut {NoStop}%
\bibitem [{\citenamefont {Baffou}\ \emph
  {et~al.}(2010{\natexlab{b}})\citenamefont {Baffou}, \citenamefont {Quidant},\
  and\ \citenamefont {Garc\'ia~de Abajo}}]{baffou2010ACS}%
  \BibitemOpen
  \bibfield  {author} {\bibinfo {author} {\bibfnamefont {G.}~\bibnamefont
  {Baffou}}, \bibinfo {author} {\bibfnamefont {R.}~\bibnamefont {Quidant}}, \
  and\ \bibinfo {author} {\bibfnamefont {F.~J.}\ \bibnamefont {Garc\'ia~de
  Abajo}},\ }\href@noop {} {\bibfield  {journal} {\bibinfo  {journal} {ACS
  nano}\ }\textbf {\bibinfo {volume} {4}},\ \bibinfo {pages} {709} (\bibinfo
  {year} {2010}{\natexlab{b}})}\BibitemShut {NoStop}%
\bibitem [{\citenamefont {Hern\'andez}\ \emph {et~al.}(2005)\citenamefont
  {Hern\'andez}, \citenamefont {Noordam},\ and\ \citenamefont
  {Robicheaux}}]{2005hernandez}%
  \BibitemOpen
  \bibfield  {author} {\bibinfo {author} {\bibfnamefont {J.~V.}\ \bibnamefont
  {Hern\'andez}}, \bibinfo {author} {\bibfnamefont {L.~D.}\ \bibnamefont
  {Noordam}}, \ and\ \bibinfo {author} {\bibfnamefont {F.}~\bibnamefont
  {Robicheaux}},\ }\href {\doibase 10.1021/jp0527352} {\bibfield  {journal}
  {\bibinfo  {journal} {J. of Phys. Chem. B}\ }\textbf {\bibinfo {volume}
  {109}},\ \bibinfo {pages} {15808} (\bibinfo {year} {2005})}\BibitemShut
  {NoStop}%
\bibitem [{\citenamefont {de~Waele}\ \emph {et~al.}(2007)\citenamefont
  {de~Waele}, \citenamefont {Koenderink},\ and\ \citenamefont
  {Polman}}]{2007dewaele}%
  \BibitemOpen
  \bibfield  {author} {\bibinfo {author} {\bibfnamefont {R.}~\bibnamefont
  {de~Waele}}, \bibinfo {author} {\bibfnamefont {A.~F.}\ \bibnamefont
  {Koenderink}}, \ and\ \bibinfo {author} {\bibfnamefont {A.}~\bibnamefont
  {Polman}},\ }\href {\doibase 10.1021/nl070807q} {\bibfield  {journal}
  {\bibinfo  {journal} {Nano Lett.}\ }\textbf {\bibinfo {volume} {7}},\
  \bibinfo {pages} {2004} (\bibinfo {year} {2007})}\BibitemShut {NoStop}%
\bibitem [{\citenamefont {Malyshev}\ \emph {et~al.}(2008)\citenamefont
  {Malyshev}, \citenamefont {Malyshev},\ and\ \citenamefont
  {Knoester}}]{2008malyshev}%
  \BibitemOpen
  \bibfield  {author} {\bibinfo {author} {\bibfnamefont {A.~V.}\ \bibnamefont
  {Malyshev}}, \bibinfo {author} {\bibfnamefont {V.~A.}\ \bibnamefont
  {Malyshev}}, \ and\ \bibinfo {author} {\bibfnamefont {J.}~\bibnamefont
  {Knoester}},\ }\href {\doibase 10.1021/nl8011962} {\bibfield  {journal}
  {\bibinfo  {journal} {Nano Lett.}\ }\textbf {\bibinfo {volume} {8}},\
  \bibinfo {pages} {2369} (\bibinfo {year} {2008})}\BibitemShut {NoStop}%
\bibitem [{\citenamefont {Agnihotri}\ \emph {et~al.}(2014)\citenamefont
  {Agnihotri}, \citenamefont {Mukherji},\ and\ \citenamefont
  {Mukherji}}]{agnihotri2014}%
  \BibitemOpen
  \bibfield  {author} {\bibinfo {author} {\bibfnamefont {S.}~\bibnamefont
  {Agnihotri}}, \bibinfo {author} {\bibfnamefont {S.}~\bibnamefont {Mukherji}},
  \ and\ \bibinfo {author} {\bibfnamefont {S.}~\bibnamefont {Mukherji}},\
  }\href@noop {} {\bibfield  {journal} {\bibinfo  {journal} {Rsc Advances}\
  }\textbf {\bibinfo {volume} {4}},\ \bibinfo {pages} {3974} (\bibinfo {year}
  {2014})}\BibitemShut {NoStop}%
\bibitem [{\citenamefont {Link}\ and\ \citenamefont
  {El-Sayed}(1999)}]{link1999}%
  \BibitemOpen
  \bibfield  {author} {\bibinfo {author} {\bibfnamefont {S.}~\bibnamefont
  {Link}}\ and\ \bibinfo {author} {\bibfnamefont {M.~A.}\ \bibnamefont
  {El-Sayed}},\ }\href@noop {} {\bibfield  {journal} {\bibinfo  {journal} {J.
  Phys. Chem. B}\ }\textbf {\bibinfo {volume} {103}},\ \bibinfo {pages} {4212}
  (\bibinfo {year} {1999})}\BibitemShut {NoStop}%
\bibitem [{\citenamefont {Bromberg}(2011)}]{bromberg2011}%
  \BibitemOpen
  \bibfield  {author} {\bibinfo {author} {\bibfnamefont {K.~A. S.~D.}\
  \bibnamefont {Bromberg}, \bibfnamefont {Sarina;~Dill}},\ }\href@noop {}
  {\emph {\bibinfo {title} {Molecular Driving Forces: Statistical
  Thermodynamics in Biology, Chemistry, Physics, and Nanoscience, 2nd
  Edition}}},\ \bibinfo {edition} {2nd}\ ed.\ (\bibinfo  {publisher} {Garland
  Science},\ \bibinfo {year} {2011})\BibitemShut {NoStop}%
\end{thebibliography}%

\end{document}